\pdfoutput=1
\documentclass[3p,times]{elsarticle}
\usepackage{amsmath,amssymb,amsthm,mathtools,bm}
\usepackage{graphicx,booktabs,array}
\usepackage[dvipsnames]{xcolor}
\usepackage[colorlinks=true,linkcolor=MidnightBlue,citecolor=MidnightBlue,urlcolor=MidnightBlue]{hyperref}
\usepackage{enumitem,microtype}
\setlist[enumerate]{topsep=3pt,parsep=4pt,itemsep=-4pt,leftmargin=*,listparindent=0pt,after=\parskip0pt\relax}
\setlist[description]{topsep=3pt,parsep=0pt,itemsep=2pt,leftmargin=1.5em}
\AtBeginDocument{\relpenalty=10000 \binoppenalty=10000 }
\journal{Physica A}

\newtheorem{definition}{Definition}
\newtheorem{lemma}{Lemma}
\newtheorem{theorem}{Theorem}
\newtheorem{proposition}{Proposition}
\theoremstyle{definition}
\newtheorem{remark}{Remark}
\newtheorem{example}{Example}

\newcommand{\E}{\mathbb E}\newcommand{\Var}{\operatorname{Var}}\newcommand{\Cov}{\operatorname{Cov}}
\newcommand{\R}{\mathbb R}\newcommand{\N}{\mathcal N}\newcommand{\one}{\bm 1}
\newcommand{\vr}{\bm r}\newcommand{\vx}{\bm x}\newcommand{\vy}{\bm y}\newcommand{\vz}{\bm z}
\newcommand{\va}{\bm a}\newcommand{\vb}{\bm b}\newcommand{\vR}{\bm R}
\newcommand{\mA}{\bm A}\newcommand{\mL}{\bm L}\newcommand{\mSigma}{\bm\Sigma}
\newcommand{\tA}{\bm{\mathcal A}}\newcommand{\tAc}{\tilde{\bm{\mathcal A}}}
\newcommand{\F}{\psi^{\times}}\newcommand{\Fl}{\psi}\newcommand{\Fcp}{\psi^{\mathrm{cp}}}
\newcommand{\AM}{\mathrm{AM}}\newcommand{\HM}{\mathrm{HM}}\newcommand{\CSV}{\mathrm{CSV}}\newcommand{\VIX}{\mathrm{VIX}}
\newcommand{\cov}{\operatorname{cov}}\newcommand{\var}{\operatorname{var}}
\newcommand{\logit}{\operatorname{logit}}\newcommand{\EMG}{\mathrm{EMG}}
\DeclareMathOperator{\tr}{tr}
\newcolumntype{L}[1]{>{\raggedright\arraybackslash}p{#1}}

\begin{document}
\begin{frontmatter}
\title{Two Faces of Financial Chaos}

\author[a]{Masoud Ataei\corref{cor1}}
\ead{masoud.ataei@utoronto.ca}
\author[b]{Shengyuan Chen}
\ead{chensy@yorku.ca}
\author[c]{Zijiang Yang}
\ead{zyang@yorku.ca}
\author[d]{Sepideh Forouzi}
\ead{sepideh.forouzi@colorado.edu}
\cortext[cor1]{Corresponding author.}
\address[a]{Department of Mathematical and Computational Sciences, University of Toronto, Mississauga, Ontario, Canada}
\address[b]{Department of Mathematics and Statistics, York University, Toronto, Ontario, Canada}
\address[c]{School of Information Technology, York University, Toronto, Ontario, Canada}
\address[d]{Department of Computer Science, University of Colorado Boulder, Boulder, Colorado, USA}

\begin{abstract}
The VIX measures the implied volatility of a stock index, and it is silent when stocks move apart while the index stays calm. The financial chaos index (FCIX) measures the realized disagreement among the returns of all pairs of stocks, and this paper studies what the FCIX records that the VIX does not. The closed form of the FCIX equals the average pairwise disagreement of returns, and its exact relation to the tensor index, its sampling error and its sensitivity to the collapse of a single stock are derived. An exact accounting splits the monthly FCIX into implied index variance, a wedge between realized and implied market variance, the realized co-movement of stocks and a form term. For the 1,000 largest U.S.\ stocks over 1990--2025, implied variance accounts for 42\% of the variance of excess chaos. Between a third and a half of the high-chaos episodes are led by dispersion and not by the VIX, and such episodes occur throughout the sample. They arise from falling co-movement, from market turmoil beyond what index options price and, in one case, from the collapse of a single stock. Months in which the news concentrates on few themes have higher implied variance and higher realized co-movement but no more dispersion, and lagged news has no robust predictive content. A regime-switching model with exponential bursts describes the eras of market chaos, and its density-forecast gains over Gaussian benchmarks stem mostly from the fat tail revealed by March 2020.
\end{abstract}


\end{frontmatter}

\section{Introduction}\label{sec:intro}

Market turmoil is usually gauged by the VIX, which measures the volatility of the S\&P~500 index that investors expect over the coming month, as priced in index options. The VIX therefore combines the expected volatility of the index with a premium for bearing variance risk~\cite{carr2009variance,bollerslev2009expected}, and its movements are routinely read as the level of fear in the stock market~\cite{bekaert2014vix}. By construction, however, the VIX describes a single portfolio. When stocks fall together, the index falls and the VIX rises. When stocks move apart, so that the gains of one part of the market offset the losses of another, the index and the VIX can remain calm although the returns of individual stocks are large and disparate. The internet-stock rout of April 1999 and the meme-stock squeeze of January 2021 are examples. The cross-sectional dispersion of returns registers such turmoil, and it has long been studied as a state of the market in its own right~\cite{lillo2000variety,campbell2001individual}. What the existing tools do not provide is a link between the two measurements which states, for a given surge in dispersion, how much of the surge the VIX registers and what the remainder consists of.

This paper studies the financial chaos index (FCIX) of Ataei et al.~\cite{ataei2021theory}, which measures the realized disagreement among the returns of all pairs of stocks, and uses it to answer that question. The central result is an accounting which splits the monthly FCIX into four observable terms, namely the implied variance of the index, a wedge between the realized variance of the broad market and the implied variance of the index, the realized co-movement of stocks, and a term due to the functional form of the index. The accounting makes it possible to classify every episode of high chaos by the term that leads it. We call an episode synchronized when implied variance leads, and fragmented when dispersion rises by more than implied variance does. To the best of our knowledge, the comparison of a dispersion index with the VIX has not been based on an accounting of this kind, which holds exactly for realized quantities and in which every term is observed.

The work sits among three classical descriptions of a stock market, which one elementary fact ties together. The variance of a portfolio equals the average variance of its constituents multiplied by a measure of their co-movement. Dispersion measures what remains of the average variance once the common movement is removed, correlation measures the co-movement, and the volatility of an index, realized or implied, measures the product. A surge in dispersion can therefore come from a rise in the volatility of all stocks, from a fall in their co-movement, or from both, and only the first of these sources is visible in the volatility of the index.

The FCIX was introduced through a tensor construction~\cite{ataei2021theory}. The returns of all assets are compared pairwise on each day, the daily comparison matrices are stacked into a third-order tensor, and the index measures the inconsistency of the slices of a rank-one approximation of that tensor, in the sense of the consistency index of Saaty~\cite{saaty1977scaling}. That work computed the index on a balanced panel of 811 stocks with complete price histories, documented its heavy-tailed distribution, and dated its structural changes. The present paper extends the earlier work in three directions. It derives the closed form of the index and adopts it as the working definition, which removes the need for a balanced panel. It establishes what the index estimates and with which error. Finally, it relates the index to implied volatility, to realized co-movement and to news.

Three literatures neighbor this work. The first treats dispersion as a state of the market. Lillo and Mantegna~\cite{lillo2000variety} introduced the dispersion of returns across an ensemble of stocks as the variety of a market and documented its behavior on crash and rally days. The cross-sectional variance of returns is a model-free measure of aggregate idiosyncratic volatility~\cite{garcia2014model}, whose rise in the late 1990s~\cite{campbell2001individual} proved to be episodic~\cite{brandt2010idiosyncratic} and is well described by occasional switches into a regime of high variance~\cite{bekaert2012aggregate}. Dispersion predicts the future volatility of the market~\cite{stivers2003firm}, it moves with the business cycle~\cite{angelidis2015stock}, and unusually low dispersion during large market moves is the classical sign of herding~\cite{christie1995following,chang2000examination}. The second literature concerns co-movement. Solnik and Roulet~\cite{solnik2000dispersion} used dispersion as an instantaneous measure of the correlation between national markets, average correlation predicts market returns~\cite{pollet2010average}, and correlation risk carries a premium in option prices~\cite{driessen2009price}. The third literature measures uncertainty from newspaper text~\cite{baker2016measuring,manela2017news}, including the trackers of equity market volatility by topic~\cite{baker2026policy} which we use. What these literatures do not offer, however, is an accounting of a dispersion index against the VIX, together with evidence on how often, and through which term, the two measures part. This paper develops it.

The results rest on classical tools, to each of which the paper adds a specific element. The theory of U-statistics of Hoeffding~\cite{hoeffding1948class} yields the sampling properties of the index, and we add a relation between the effective and the realized kurtosis which makes the sampling error computable from the data. The relation between dispersion and correlation of Solnik and Roulet~\cite{solnik2000dispersion} holds in expectation, and we add its realized counterpart, which holds for every sample. Markov-switching models~\cite{hamilton1989new} with exponentially modified Gaussian innovations, under which chaos follows the modified lognormal power-law distribution~\cite{basu2015mlp} within a regime, describe the eras and the bursts of chaos, and we add the tail index of chaos within a regime. Inference on many news effects uses the horseshoe prior~\cite{carvalho2010horseshoe} and a wild bootstrap~\cite{goncalves2004bootstrapping}, since Wald tests with autocorrelation-robust covariance estimators reject too often in finite samples~\cite{kiefer2005new,lazarus2018har}.

The first group of results concerns the index. The closed form of the tensor index is the ratio of the arithmetic to the harmonic mean of the gross returns, and it coincides with the tensor index exactly when the consensus weights all assets equally. An explicit factor gives the difference otherwise. The closed form, its logarithmic counterpart and the cross-sectional variance of returns are three averages of pairwise disagreements, which differ in how strongly they weight extreme pairs. The index is an unbiased and asymptotically normal estimator of the cross-sectional variance under a finite fourth moment, and we quantify the months in which the collapse of a single stock dominates it.

The second group of results concerns the two faces. The chaos accounting holds exactly in every month. Applied to the 1,000 largest U.S.\ stocks from 1990 to 2025, it shows that implied variance accounts for 42\% of the variance of excess chaos. Of 38 annotated episodes, 13 are fragmented, as are 17 of the 35 episodes selected by a mechanical rule and 40\% of all months of high chaos, and the share is similar in the two halves of the sample. Fragmented episodes are of the same order of magnitude as synchronized ones, and they are led by falling co-movement, by turmoil of the broad market in excess of what the index options price, or, in December 2021, by the collapse of one stock. The ratio of dispersion to implied variance yields a chaos-implied correlation which tracks the level of option-implied correlation, although its monthly changes add little to those of the VIX.

The third group of results concerns news and dynamics. Months of intense news have more chaos of both kinds and higher co-movement. Months in which the news concentrates on few themes have higher implied variance and higher realized co-movement, and no more dispersion. Lagged news has no predictive content for chaos once the tests are made robust. A three-regime model with exponential bursts describes calm, turbulent and highly chaotic eras, although the Bayesian information criterion prefers a single regime once long-memory terms are included. The bursts amount to right-skewed and fat-tailed innovations, and three quarters of their gain in density forecasts over a Gaussian benchmark are obtained by a symmetric fat-tailed benchmark and stem from March 2020.

The generality of the results is as follows. The closed form, the representations and the chaos accounting are algebraic. They hold for every cross-section of positive gross returns and every month, with no model of returns. The sampling results assume that the idiosyncratic shocks are independent across assets and have finite fourth moments, and we state what fails otherwise. The empirical results hold for monthly averages of daily indexes of large U.S.\ stocks, and the associations with news are contemporaneous descriptions with no causal interpretation.

The application is the monitoring of market stress. The FCIX and its accounting against the VIX identify the episodes of turmoil which implied volatility does not register, and the chaos-implied correlation can be computed for any group of stocks and any period, with no options on the individual stocks.

The remainder of the paper is organized as follows. Section~\ref{sec:fcix} develops the financial chaos index, with its tensor definition, its closed form, its representations, its sampling properties, and its relation to co-movement and to implied variance, which gives the chaos accounting and the definition of the two faces. Section~\ref{sec:faces} describes the data and documents the two faces of chaos in the episodes of 1990--2025. Section~\ref{sec:news} relates the two faces to the intensity, the concentration and the themes of news. Section~\ref{sec:eras} describes the eras and the bursts of chaos and evaluates density forecasts. Section~\ref{sec:conclusion} concludes. The longer proofs are collected in \ref{app:proofs}.

\section{Financial Chaos Index}\label{sec:fcix}

This section recalls the tensor definition of the financial chaos index, derives its closed form together with the exact relation between the two, and adopts the closed form as the working definition. It then gives the representations which show what the index measures, and the sampling properties which show how accurately the index measures it. Finally, it relates the index to the co-movement of stocks and to implied variance, which yields an accounting of the monthly index in four observed terms and the definition of the two faces.

Throughout the paper, scalars, vectors and matrices are denoted by lowercase lightface, lowercase boldface and uppercase boldface letters, respectively, and boldface calligraphic letters denote tensors. All vectors are column vectors, $\one$ is the vector of ones, and $1/\vx$ is the entrywise reciprocal of a positive vector $\vx$. For vectors $\bm u,\bm w\in\R^N$, we write
\begin{equation*}
\cov(\bm u,\bm w)=\frac1N\sum_{i=1}^N(u_i-\bar u)(w_i-\bar w),\qquad \var(\bm u)=\cov(\bm u,\bm u),
\end{equation*}
for the cross-sectional covariance and variance, where a bar denotes the cross-sectional mean. Trading days are indexed by $d=1,\dots,n$ and months by $t=1,\dots,T$, the set of trading days of month $t$ is also denoted by $t$, and $n_t$ is their number.

Consider $N\ge2$ assets, and let $C_{i,d}$ be the adjusted closing price of asset $i$ on day $d$. The gross return, the log return and the simple return of asset $i$ on day $d$ are
\begin{equation*}
r_{i,d}=\frac{C_{i,d}}{C_{i,d-1}},\qquad \ell_{i,d}=\log r_{i,d},\qquad R_{i,d}=r_{i,d}-1,
\end{equation*}
respectively, and they are collected in the vectors $\vr_d\in\R^N_{>0}$, $\bm\ell_d\in\R^N$ and $\vR_d\in\R^N$. The construction of Ataei et al.~\cite{ataei2021theory} starts from the comparison of every asset with every other asset on a given day.

\begin{definition}[Pairwise comparison matrix]\label{def:pcm}
The pairwise comparison matrix of day $d$ is the matrix $\mA_d\in\R^{N\times N}_{>0}$ with elements
\begin{equation*}
a_{ij,d}=\frac{r_{j,d}}{r_{i,d}},\qquad i,j=1,\dots,N,
\end{equation*}
that is, $\mA_d=(1/\vr_d)\,\vr_d^\top$.
\end{definition}

The element $a_{ij,d}$ exceeds one if asset $j$ outperformed asset $i$ on day $d$ and falls below one otherwise. Its logarithm,
\begin{equation}\label{eq:logadv}
L_{ij,d}=\log a_{ij,d}=\ell_{j,d}-\ell_{i,d},
\end{equation}
is called the \emph{pairwise log-advantage} of asset $j$ over asset $i$. Matrices of ratios have the following structure, which was established for pairwise comparative judgments by Saaty~\cite{saaty1977scaling}.

\begin{lemma}[Consistency]\label{lem:slice}
Let $\mA_d$ be the pairwise comparison matrix of day $d$. Then, the following hold.
\begin{enumerate}[label=\normalfont(\roman*)]
\item The matrix $\mA_d$ is reciprocal, that is,
\begin{equation*}
a_{ii,d}=1,\qquad a_{ij,d}\,a_{ji,d}=1,\qquad i,j=1,\dots,N.
\end{equation*}
\item The matrix $\mA_d$ is consistent, that is,
\begin{equation*}
a_{ik,d}\,a_{kj,d}=a_{ij,d},\qquad i,j,k=1,\dots,N.
\end{equation*}
\item The matrix $\mA_d$ has rank one, and its only nonzero eigenvalue is
\begin{equation}\label{eq:lmaxN}
\lambda_{\max}(\mA_d)=N.
\end{equation}
\end{enumerate}
\end{lemma}

\begin{proof}
The three parts are proved in order.
\begin{enumerate}[label=\normalfont(\roman*)]
\item By Definition~\ref{def:pcm}, $a_{ii,d}=r_{i,d}/r_{i,d}=1$ and $a_{ij,d}\,a_{ji,d}=(r_{j,d}/r_{i,d})(r_{i,d}/r_{j,d})=1$.
\item Similarly, $a_{ik,d}\,a_{kj,d}=(r_{k,d}/r_{i,d})(r_{j,d}/r_{k,d})=a_{ij,d}$.
\item The matrix $\mA_d=(1/\vr_d)\,\vr_d^\top$ is the outer product of two nonzero vectors and has rank one. A rank-one matrix $\bm u\bm w^\top$ has the single nonzero eigenvalue $\bm w^\top\bm u$, with eigenvector $\bm u$, and here $\vr_d^\top(1/\vr_d)=N$.\qedhere
\end{enumerate}
\end{proof}

Lemma~\ref{lem:slice} states that the comparisons of a single day are perfectly coherent, and that the largest eigenvalue of every comparison matrix equals $N$ whatever the returns are. A departure from coherence can therefore arise only when the comparisons of a day are confronted with a pattern which the day does not determine. In the tensor construction, this pattern is a consensus extracted from all days. Recall that a third-order tensor $\bm{\mathcal X}\in\R^{N\times N\times n}$ has the frontal slices $\bm X_d\in\R^{N\times N}$, $d=1,\dots,n$, that its Frobenius norm $\|\bm{\mathcal X}\|_F$ is the square root of the sum of its squared elements, and that it has rank one if it is the outer product $\vz\circ\vx\circ\vy$ of three vectors, in which case its $d$th frontal slice equals $z_d\,\vx\vy^\top$.

\begin{definition}[Pairwise comparison tensor and consensus tensor]\label{def:pct}
The pairwise comparison tensor is the third-order tensor $\tA\in\R^{N\times N\times n}_{>0}$ whose $d$th frontal slice is the pairwise comparison matrix $\mA_d$. The consensus tensor of $\tA$ is the rank-one tensor $\tAc=\tilde\vz\circ\tilde\vx\circ\tilde\vy$, with priority vectors $\tilde\vx,\tilde\vy\in\R^N_{>0}$ and scale vector $\tilde\vz\in\R^n$, which solves
\begin{equation}\label{eq:cpd}
\min_{\tilde\vx,\tilde\vy,\tilde\vz}\big\|\tA-\tilde\vz\circ\tilde\vx\circ\tilde\vy\big\|_F\qquad\text{subject to}\qquad \tilde z_d\,\tilde\vx\tilde\vy^\top>\bm 0,\quad d=1,\dots,n.
\end{equation}
The $d$th frontal slice of the consensus tensor is $\tilde\mA_d=\tilde z_d\,\tilde\vx\tilde\vy^\top$.
\end{definition}

The consensus tensor describes a hypothetical market in which the comparisons of every day are generated by the same pair of priority vectors, and the scale $\tilde z_d$ absorbs the departure of day $d$ from this common pattern. Unlike the slices of $\tA$, the slices of the consensus tensor need not be consistent, and the index of Ataei et al.~\cite{ataei2021theory} measures their inconsistency.

\begin{definition}[Tensor index]\label{def:tensorindex}
The tensor index of day $d$ is
\begin{equation}\label{eq:fcixdef}
\Fcp_d=\frac{\lambda_{\max}\big(\tilde\mA_d\big)-N}{N-1},
\end{equation}
where $\lambda_{\max}(\tilde\mA_d)$ is the largest eigenvalue of the $d$th frontal slice of the consensus tensor.
\end{definition}

By the eigenvalue~\eqref{eq:lmaxN}, the numerator of the tensor index~\eqref{eq:fcixdef} is the excess of the largest eigenvalue of the consensus slice over that of a consistent matrix, and the division by $N-1$ follows the consistency index of Saaty~\cite{saaty1977scaling}. Definition~\ref{def:tensorindex} seems to require the solution of the nonconvex problem~\eqref{eq:cpd}. The next lemma shows that the part of the solution which varies with the day is available in closed form.

\begin{lemma}[Consensus scale]\label{lem:scale}
Let $\tilde\vx,\tilde\vy\in\R^N_{>0}$ be fixed. Then, the minimizer of $\|\mA_d-z\,\tilde\vx\tilde\vy^\top\|_F$ over $z\in\R$ and the largest eigenvalue of the slice $z^\star_d\,\tilde\vx\tilde\vy^\top$ are
\begin{equation}\label{eq:zstar}
z^\star_d=\frac{\big(\tilde\vx^\top(1/\vr_d)\big)\big(\tilde\vy^\top\vr_d\big)}{\|\tilde\vx\|^2\,\|\tilde\vy\|^2}>0,\qquad \lambda_{\max}\big(z^\star_d\,\tilde\vx\tilde\vy^\top\big)=z^\star_d\,\tilde\vy^\top\tilde\vx,
\end{equation}
respectively.
\end{lemma}

\begin{proof}
The squared objective is a convex quadratic function of $z$, namely
\begin{equation*}
\big\|\mA_d-z\,\tilde\vx\tilde\vy^\top\big\|_F^2=\|\mA_d\|_F^2-2z\,\big\langle\mA_d,\tilde\vx\tilde\vy^\top\big\rangle+z^2\|\tilde\vx\|^2\|\tilde\vy\|^2,
\end{equation*}
where $\langle\cdot,\cdot\rangle$ is the Frobenius inner product, and it is minimized at the ratio of $\langle\mA_d,\tilde\vx\tilde\vy^\top\rangle$ to $\|\tilde\vx\|^2\|\tilde\vy\|^2$. By Definition~\ref{def:pcm},
\begin{equation*}
\big\langle\mA_d,\tilde\vx\tilde\vy^\top\big\rangle=\sum_{i,j=1}^N\tilde x_i\,\tilde y_j\,\frac{r_{j,d}}{r_{i,d}}=\big(\tilde\vx^\top(1/\vr_d)\big)\big(\tilde\vy^\top\vr_d\big)>0,
\end{equation*}
which gives the first formula. The slice $z^\star_d\,\tilde\vx\tilde\vy^\top$ has rank one, and its only nonzero eigenvalue is $z^\star_d\,\tilde\vy^\top\tilde\vx$ by the argument of Lemma~\ref{lem:slice}(iii).
\end{proof}

Lemma~\ref{lem:scale} shows that the positivity constraint of the problem~\eqref{eq:cpd} is inactive, and that the returns of day $d$ enter the tensor index only through the scale $z^\star_d$, which is the product of a weighted average of the gross returns and a weighted average of their reciprocals. The priority vectors, in contrast, are common to all days. The next theorem expresses the tensor index through the arithmetic and harmonic means of the gross returns of the day, for arbitrary priority vectors, and identifies the case in which the priority vectors drop out.

\begin{theorem}[Closed form]\label{thm:closed}
Let $\tilde\vx,\tilde\vy\in\R^N_{>0}$ be priority vectors with means $\bar x$ and $\bar y$, let $\va=\tilde\vx/\bar x-\one$ and $\vb=\tilde\vy/\bar y-\one$ be their relative deviations from uniformity, and let $\Fcp_d$ be the tensor index~\eqref{eq:fcixdef} with the consensus scale~\eqref{eq:zstar}. Let further
\begin{equation*}
\AM_d=\frac1N\sum_{i=1}^Nr_{i,d},\qquad \HM_d=\bigg(\frac1N\sum_{i=1}^N\frac{1}{r_{i,d}}\bigg)^{-1}
\end{equation*}
be the arithmetic and harmonic means of the gross returns of day $d$, respectively. Then, the following hold.
\begin{enumerate}[label=\normalfont(\roman*)]
\item For every day $d$, the tensor index satisfies
\begin{equation}\label{eq:theta}
\Fcp_d=\frac{N}{N-1}\bigg(\frac{\AM_d}{\HM_d}\,\Theta_d-1\bigg),\qquad
\Theta_d=\frac{\big(1+\HM_d\cov(\va,1/\vr_d)\big)\big(1+\cov(\vb,\vr_d)/\AM_d\big)\big(1+\cov(\va,\vb)\big)}{\big(1+\var(\va)\big)\big(1+\var(\vb)\big)}.
\end{equation}
\item If the priority vectors are uniform, that is, $\va=\vb=\bm 0$, then $\Theta_d=1$ for every day $d$, and
\begin{equation}\label{eq:fcix}
\Fcp_d=\frac{N}{N-1}\bigg(\frac{\AM_d}{\HM_d}-1\bigg).
\end{equation}
\item On a day on which all assets earn the same return, the right-hand side of the closed form~\eqref{eq:fcix} vanishes, whereas
\begin{equation*}
\Fcp_d\le0,
\end{equation*}
with equality if and only if $\va=\vb=\bm 0$.
\end{enumerate}
\end{theorem}

The proof is given in \ref{app:proofs}. Theorem~\ref{thm:closed} separates what the tensor index owes to the returns of the day from what it owes to the consensus. The returns enter through the ratio of their arithmetic to their harmonic mean, and the consensus enters through the factor $\Theta_d$, which depends on the day only through the covariances of the priority deviations with the returns and with their reciprocals. The arithmetic mean of positive numbers is never smaller than their harmonic mean, with equality if and only if the numbers coincide. Hence, the closed form~\eqref{eq:fcix} is nonnegative, vanishes exactly when all assets earn the same return, and grows as the returns spread out. Part~(iii) shows that the tensor index has this property only for uniform priority vectors. The closed form~\eqref{eq:fcix} has two further advantages. It is a function of the cross-section of the day alone, so that it can be computed for a universe of assets which changes from day to day, whereas the tensor requires a balanced panel. Moreover, it does not depend on the sample through fitted priority vectors, so that the index of a given day is not revised when later days are added. We therefore take the closed form as the definition of the index, together with two companions which replace the ratio of means by a variance.

\begin{definition}[Financial chaos index]\label{def:fcix}
The financial chaos index of day $d$ in its multiplicative form, its logarithmic form and its simple-return form is
\begin{equation}\label{eq:forms}
\F_d=\frac{N}{N-1}\bigg(\frac{\AM_d}{\HM_d}-1\bigg),\qquad \Fl_d=\frac1N\sum_{i=1}^N\big(\ell_{i,d}-\bar\ell_d\big)^2,\qquad \CSV_d=\frac1N\sum_{i=1}^N\big(R_{i,d}-\bar R_d\big)^2,
\end{equation}
respectively, where $\bar\ell_d$ and $\bar R_d$ are the cross-sectional means of the log returns and of the simple returns.
\end{definition}

\begin{remark}[Uniform priority vectors]\label{rem:uniform}
The hypothesis of Theorem~\ref{thm:closed}(ii) is exact uniformity, and Theorem~\ref{thm:closed}(i) bounds the effect of a departure from it. Let $s_a$ and $s_b$ be the cross-sectional standard deviations of $\va$ and $\vb$, and let $\kappa_d$ and $\kappa'_d$ be the cross-sectional coefficients of variation of $1/\vr_d$ and $\vr_d$. By the Cauchy--Schwarz inequality applied to each covariance of the factor $\Theta_d$,
\begin{equation*}
\frac{(1-s_a\kappa_d)(1-s_b\kappa'_d)(1-s_as_b)}{(1+s_a^2)(1+s_b^2)}\le\Theta_d\le(1+s_a\kappa_d)(1+s_b\kappa'_d)(1+s_as_b).
\end{equation*}
Since $\kappa_d^2$ and $\kappa_d'^2$ are close to $\Fl_d$, the relative difference between the tensor index and the closed form is at most of order $(s_a+s_b)/\sqrt{\Fl_d}$, and this order is attained only if the priority deviations are collinear with the returns of the day. At a stationary point of the problem~\eqref{eq:cpd}, the first-order conditions make $\tilde\vx$ a positively weighted average over days of the vectors $1/\vr_d$ and $\tilde\vy$ a positively weighted average of the vectors $\vr_d$, so that the priority deviations are of the order of the average daily return of an asset in excess of the market. In our earlier computations on a Compustat-based panel of U.S.\ equities, the fitted priority vectors deviated from uniformity by at most $0.003$, and the tensor index had a correlation of $0.9999$ with the closed form.
\end{remark}

The three forms of Definition~\ref{def:fcix} are closely related measures of the disagreement among returns. The next theorem makes the relation precise. It shows that each form is an average over pairs of assets, identifies the multiplicative form with a classical consistency index and the logarithmic form with an energy, and quantifies the gaps between the forms.

\begin{theorem}[Representations]\label{thm:ustat}
Let $\ell_i$, $R_i$ and $r_i$ be the log returns, the simple returns and the gross returns of $N$ assets on a given day, whose subscript is dropped, and let $\F$, $\Fl$ and $\CSV$ be the three forms~\eqref{eq:forms} of the index. Let $\mL=[L_{ij}]$ be the matrix of the pairwise log-advantages~\eqref{eq:logadv}, and let $e_i=r_i/\AM-1$ be the relative deviations of the gross returns from their arithmetic mean, with moments $m_k=N^{-1}\sum_{i=1}^Ne_i^k$. Then, the following hold.
\begin{enumerate}[label=\normalfont(\roman*)]
\item Each form is an average over all pairs of assets, namely
\begin{equation*}
\F=\binom N2^{-1}\sum_{i<j}\big\{\cosh(\ell_i-\ell_j)-1\big\},\qquad \frac{N\,\Fl}{N-1}=\binom N2^{-1}\sum_{i<j}\frac{(\ell_i-\ell_j)^2}{2},\qquad \frac{N\,\CSV}{N-1}=\binom N2^{-1}\sum_{i<j}\frac{(R_i-R_j)^2}{2}.
\end{equation*}
\item The multiplicative form is the consistency index of Saaty evaluated at the uniform priority vector, that is,
\begin{equation*}
\F=\frac{\mathcal S-N}{N-1},\qquad \mathcal S=\frac1N\sum_{i,j=1}^Na_{ij}.
\end{equation*}
\item The matrix $\mL$ is skew-symmetric of rank at most two, its two largest singular values are both equal to $\sqrt N\,\|\bm\ell-\bar\ell\one\|$, and
\begin{equation*}
\Fl=\frac{\|\mL\|_F^2}{2N^2}.
\end{equation*}
\item The gap between the multiplicative and the logarithmic form is nonnegative and is given by the even moments of the pairwise log-advantages, namely
\begin{equation}\label{eq:gap}
\frac{N-1}{N}\,\F-\Fl=\sum_{k\ge2}\frac{M_{2k}}{(2k)!}\ge0,\qquad M_{2k}=\frac{1}{N^2}\sum_{i,j=1}^NL_{ij}^{2k}.
\end{equation}
The leading term of the series equals $\Fl^2(\mathcal K+3)/12$, where $\mathcal K$ is the cross-sectional kurtosis of the log returns.
\item The multiplicative and the simple-return form are the weighted and the unweighted second moment of the relative deviations, namely
\begin{equation*}
\frac{N-1}{N}\,\F=\frac1N\sum_{i=1}^N\frac{e_i^2}{1+e_i},\qquad \CSV=\AM^2\,m_2,
\end{equation*}
and the logarithmic form satisfies, as $\max_i|e_i|\to0$ and uniformly in $N$,
\begin{equation}\label{eq:logexp}
\Fl=m_2-m_3+\frac{11}{12}\,m_4-\frac14\,m_2^2+O\Big(\max_i|e_i|^5\Big).
\end{equation}
\end{enumerate}
\end{theorem}

The proof is given in \ref{app:proofs}. Part~(i) states that the index is the average, over all $N(N-1)/2$ pairs of assets, of a convex and even function of the difference between two returns. The three forms differ in this function only. The multiplicative form weights the log-advantage of a pair exponentially, the logarithmic form weights it quadratically, and the simple-return form weights the difference of simple returns quadratically. This reading explains why a measure of dispersion, a statistic of herding~\cite{christie1995following} and an index of consistency coincide. The representation of part~(ii) is the consistency index of Saaty~\cite{saaty1977scaling}, and the present result adds that the multiplicative index is its value at the consensus according to which all assets perform alike. Parts~(iv) and~(v) show how the forms part. The multiplicative form exceeds the logarithmic form by a term which grows with the cross-sectional kurtosis. By the expansion~\eqref{eq:logexp}, the logarithmic and the multiplicative form both carry the third moment $m_3$ with a negative sign, whereas the simple-return form does not carry it. A large negative $m_3$ arises when one stock loses most of its value. Indeed, a stock with a large simple return $R$ contributes approximately $\{\log(1+R)\}^2/N$ to the logarithmic form and $R^2/N$ to the simple-return form, and the ratio of the two contributions at $R=-0.9$ is $(\log 0.1/0.9)^2=6.546$. The logarithmic form is thus more sensitive than the simple-return form to the collapse of a single stock, and the multiplicative form is more sensitive still. In the empirical analysis, we work with the logarithmic form, which is the cross-sectional variance of the log returns, and we report its gap to the simple-return form explicitly.

We next study the index as an estimator. At the monthly frequency, we work with the logarithm of the monthly mean of the daily index, that is,
\begin{equation}\label{eq:monthly}
y_t=\log\bigg(\frac{1}{n_t}\sum_{d\in t}\Fl_d\bigg).
\end{equation}
The next theorem states what the daily index and the monthly index~\eqref{eq:monthly} estimate, and with which error. Its first two parts treat a cross-section which is independent and identically distributed given the common shocks of the day, and its last two parts allow for assets with different conditional means and variances. The statements are given for the logarithmic form. They hold verbatim for the simple-return form, with the log returns replaced by the simple returns, since both forms are cross-sectional variances.

\begin{theorem}[Sampling properties]\label{thm:clt}
Let $\Fl_d$ be the logarithmic index of $N$ assets on day $d$, let $y_t$ be the monthly index~\eqref{eq:monthly}, and let $\mathcal G$ be a $\sigma$-field. Then, the following hold.
\begin{enumerate}[label=\normalfont(\roman*)]
\item If, conditionally on $\mathcal G$, the log returns $\ell_{1,d},\dots,\ell_{N,d}$ are independent and identically distributed with variance $D_d$, then the rescaled index is unbiased for every $N\ge2$, that is,
\begin{equation*}
\E\Big[\frac{N}{N-1}\,\Fl_d\,\Big|\,\mathcal G\Big]=D_d.
\end{equation*}
\item If, in addition, the conditional kurtosis $\mathcal K^\ast_d$ of the log returns is finite, then, conditionally on $\mathcal G$ and as $N\to\infty$, the convergence in distribution
\begin{equation}\label{eq:clt}
\sqrt N\,\big(\log\Fl_d-\log D_d\big)\Rightarrow\N\big(0,\mathcal K^\ast_d-1\big)
\end{equation}
holds.
\item Suppose instead that, conditionally on $\mathcal G$ and for the days $d$ of month $t$, the log returns satisfy $\ell_{i,d}=\theta_{i,d}+\varepsilon_{i,d}$, where the means $\theta_{i,d}$ are $\mathcal G$-measurable and uniformly bounded, and the shocks $\varepsilon_{i,d}$ are independent across assets and days, with mean zero, variances $h_{i,d}$ and uniformly bounded fourth moments. Let
\begin{equation*}
s_{i,d}=\big(\theta_{i,d}-\bar\theta_d\big)^2+h_{i,d},\qquad D_d=\frac1N\sum_{i=1}^Ns_{i,d},\qquad \mathcal K^\circ_d=1+\frac{N\Var(\Fl_d\mid\mathcal G)}{D_d^2},
\end{equation*}
and suppose that $D_d$ is bounded away from zero. Then, the monthly index satisfies
\begin{equation}\label{eq:measerr}
y_t=\log\bigg(\frac{1}{n_t}\sum_{d\in t}D_d\bigg)+U_t+O_p\big(N^{-1}\big),\qquad
\Var(U_t\mid\mathcal G)=\frac{\sum_{d\in t}D_d^2\,(\mathcal K^\circ_d-1)}{N\big(\sum_{d\in t}D_d\big)^2}\le\frac{\max_{d\in t}\mathcal K^\circ_d-1}{N\,n^{\mathrm{eff}}_t},
\end{equation}
where $U_t$ is the ratio of $\sum_{d\in t}(\Fl_d-D_d)$ to $\sum_{d\in t}D_d$, and $n^{\mathrm{eff}}_t$ is the ratio of $(\sum_{d\in t}D_d)^2$ to $\sum_{d\in t}D_d^2$, which lies between $1$ and $n_t$.
\item In the setting of part~(iii), the effective kurtosis $\mathcal K^\circ_d$ satisfies
\begin{equation}\label{eq:keff}
\mathcal K^\circ_d=\frac{\bar\mu_{4,d}}{D_d^2}-\frac{\var(\bm s_d)}{D_d^2}+O\big(N^{-1}\big),\qquad \bar\mu_{4,d}=\frac1N\sum_{i=1}^N\E\Big[\big(\theta_{i,d}-\bar\theta_d+\varepsilon_{i,d}\big)^4\,\Big|\,\mathcal G\Big],
\end{equation}
where $\bm s_d=(s_{1,d},\dots,s_{N,d})^\top$.
\end{enumerate}
\end{theorem}

The proof is given in \ref{app:proofs}. Parts~(i) and~(ii) are instances of the theory of $U$-statistics of Hoeffding~\cite{hoeffding1948class}, applied to the representation of Theorem~\ref{thm:ustat}(i). They state that the index of a single day is an unbiased and asymptotically lognormal estimate of the conditional cross-sectional variance $D_d$, which we call the \emph{dispersion variance}, with a precision governed by the kurtosis of the cross-section. Part~(iii) carries the result over to the monthly index for assets with different means and variances, and shows that averaging over the days of a month divides the error variance by the effective number of days $n^{\mathrm{eff}}_t$. The present result adds part~(iv), which makes the bound~\eqref{eq:measerr} computable. The first term on the right-hand side of the relation~\eqref{eq:keff} is the population counterpart of the realized cross-sectional kurtosis, and the second term is the squared coefficient of variation of the second moments $s_{i,d}$ of the individual assets. The realized kurtosis therefore overstates the effective kurtosis, and using it in the bound~\eqref{eq:measerr} is conservative.

\begin{remark}[Fourth moment]\label{rem:fourth}
The finite kurtosis of Theorem~\ref{thm:clt}(ii) is essential for the rate $N^{-1/2}$. Suppose that the cross-section of log returns has a power-law tail with exponent $\vartheta\in(2,4)$. The index $\Fl_d$ still converges to $D_d$ by the law of large numbers, since the variance is finite. The squared deviations, however, have a tail with exponent $\vartheta/2\in(1,2)$ and an infinite variance, so that their average lies in the domain of attraction of a stable law, and the error $\Fl_d-D_d$ is of order $N^{-(1-2/\vartheta)}$ up to a slowly varying factor~\cite{embrechts1997modelling}. An exponent close to three has been reported for the returns of individual stocks~\cite{plerou1999scaling,gabaix2003theory}. If the cross-section of a day has this exponent, the error of the daily index is of order $N^{-1/3}$, and the normal limit~\eqref{eq:clt} does not apply.
\end{remark}

\begin{example}\label{ex:onefactor-index}
Let the returns of the $N$ assets be generated by a one-factor market. That is, the simple returns of a given day are
\begin{equation}\label{eq:onefactor}
R_i=b_i\,F+\varepsilon_i,\qquad i=1,\dots,N,
\end{equation}
where the factor $F$ has mean zero and variance $\sigma_F^2$, the loadings $b_i$ are independent draws from $\N(\bar b,s_b^2)$, and the idiosyncratic shocks $\varepsilon_i$ are independent draws from $\N(0,\sigma_\varepsilon^2)$, independent of the loadings and of the factor. We take $N=10^3$, $\bar b=1$, $s_b=0.3$, $\sigma_F=0.01$ and $\sigma_\varepsilon=0.02$ throughout this section. Conditionally on the factor, the returns are independent and Gaussian with the dispersion variance
\begin{equation*}
D=s_b^2F^2+\sigma_\varepsilon^2,
\end{equation*}
so that their kurtosis equals $3$. By Theorem~\ref{thm:clt}(i) and~(ii), applied to the simple-return form, the rescaled index is unbiased for $D$, and the asymptotic standard deviation of its logarithm is $\sqrt{2/N}=0.045$. For a month of $21$ days with equal dispersion variances, the bound~\eqref{eq:measerr} gives the standard deviation $\sqrt{2/(21N)}=0.010$ for the monthly index. The dispersion variance depends on the factor through the dispersion of the loadings. On a day on which the factor moves by two standard deviations, the dispersion variance is $1.09\,\sigma_\varepsilon^2$, against $\sigma_\varepsilon^2$ on a day on which the factor does not move.
\end{example}

The index thus estimates the dispersion variance of the cross-section. We now relate that dispersion to the co-movement of stocks, first in expectation and then for realized returns, and bring implied variance into the relation. In doing so, the index is compared with the VIX, and it is useful to recall the nature of the two measures. The FCIX is computed from returns which have occurred and measures how much the assets moved relative to one another. The squared VIX is a risk-neutral expectation of the variance of the S\&P~500 index over the next 30 days, extracted from option prices, and it contains a variance risk premium~\cite{carr2009variance}.

Let $\vR_d$ be the vector of the simple returns of the $N$ assets on day $d$, with conditional mean $\bm\mu_d$ and conditional covariance matrix $\mSigma_d$ given the information $\mathcal F_{d-1}$ available on day $d-1$. The average variance of the assets, the variance of the equal-weighted portfolio, the co-movement ratio and the cross-sectional variance of the expected returns are
\begin{equation}\label{eq:comove}
\bar\sigma^2_d=\frac{\tr\mSigma_d}{N},\qquad \sigma^2_{\mathrm{EW},d}=\frac{\one^\top\mSigma_d\one}{N^2},\qquad \bar\rho_d=\frac{\sigma^2_{\mathrm{EW},d}}{\bar\sigma^2_d},\qquad V_d=\var(\bm\mu_d),
\end{equation}
respectively. The co-movement ratio $\bar\rho_d$ measures the extent to which the assets move as one. When all assets have the same variance, it is an affine function of the average of the pairwise correlations $\rho_{ij,d}$, namely
\begin{equation*}
\bar\rho_d=\frac1N+\frac{2}{N^2}\sum_{i<j}\rho_{ij,d}.
\end{equation*}
The realized counterpart of the co-movement ratio uses the equal-weighted return $\bar R_d$ and the simple-return form $\CSV_d$ of the index. Since the average squared return of a day equals $\bar R_d^{\,2}+\CSV_d$, we define the \emph{realized co-movement ratio} of month $t$ as the share of the common component in the sum of squared returns of the month, that is,
\begin{equation}\label{eq:rhotilde}
\tilde\rho_t=\frac{\sum_{d\in t}\bar R_d^{\,2}}{\sum_{d\in t}\big(\bar R_d^{\,2}+\CSV_d\big)}.
\end{equation}
The next theorem relates dispersion to co-movement in expectation, shows that the same relation holds exactly for the realized quantities of every month, and states what the realized co-movement ratio estimates.

\begin{theorem}[Dispersion and co-movement]\label{thm:dci}
Let $\vR_d$ be the vector of the simple returns of $N$ assets on day $d$, with conditional mean $\bm\mu_d$ and conditional covariance matrix $\mSigma_d$ given $\mathcal F_{d-1}$, let $\bar R_d$ and $\CSV_d$ be the cross-sectional mean and variance of $\vR_d$, and let $\bar\sigma^2_d$, $\sigma^2_{\mathrm{EW},d}$, $\bar\rho_d$ and $V_d$ be the quantities~\eqref{eq:comove}. Then, the following hold.
\begin{enumerate}[label=\normalfont(\roman*)]
\item The co-movement ratio satisfies $0\le\bar\rho_d\le1$, and the expected dispersion is
\begin{equation*}
\E\big[\CSV_d\mid\mathcal F_{d-1}\big]=\bar\sigma^2_d\,(1-\bar\rho_d)+V_d.
\end{equation*}
\item For every month $t$ and every realization of the returns such that $\sum_{d\in t}\bar R_d^{\,2}$ and $\sum_{d\in t}\CSV_d$ are positive, the realized co-movement ratio~\eqref{eq:rhotilde} satisfies
\begin{equation}\label{eq:realized}
\log\bigg(\frac{1}{n_t}\sum_{d\in t}\CSV_d\bigg)=\log\bigg(\frac{1}{n_t}\sum_{d\in t}\bar R_d^{\,2}\bigg)-\logit\tilde\rho_t.
\end{equation}
\item If $\bm\mu_d=\bm 0$, then
\begin{equation*}
\E\big[\bar R_d^{\,2}\mid\mathcal F_{d-1}\big]=\sigma^2_{\mathrm{EW},d},\qquad \E\big[\bar R_d^{\,2}+\CSV_d\mid\mathcal F_{d-1}\big]=\bar\sigma^2_d.
\end{equation*}
\end{enumerate}
\end{theorem}

The proof is given in \ref{app:proofs}. Part~(i) is the relation which Solnik and Roulet~\cite{solnik2000dispersion} used to estimate the correlation of national stock markets from the dispersion of their returns. It states that the cross-section disperses either because the assets become more volatile or because they cease to move together. The relation holds in expectation, and it leaves the dispersion of expected returns as a remainder. The present result adds part~(ii), in which the same relation holds for the realized quantities of every month with no expectation, no model and no remainder, once co-movement is measured by the ratio~\eqref{eq:rhotilde}. Part~(iii) states what this ratio estimates. When expected daily returns are negligible, its numerator and its denominator are sums of conditionally unbiased estimates of the numerator and of the denominator of the co-movement ratio $\bar\rho_d$, so that $\tilde\rho_t$ estimates the co-movement ratio of the month with each day weighted by its average variance.

Implied variance enters as follows. The VIX is quoted as an annualized volatility in percent, so that the implied variance of the S\&P~500 per trading day is
\begin{equation*}
\sigma^2_{Q,d}=\frac{\VIX_d^2}{252\times10^4}.
\end{equation*}
Along with the monthly index $y_t$ of equation~\eqref{eq:monthly}, we consider the logarithms of the monthly means of the simple-return form, of the squared equal-weighted return and of the implied variance, namely
\begin{equation}\label{eq:monthlyseries}
y^{s}_t=\log\bigg(\frac{1}{n_t}\sum_{d\in t}\CSV_d\bigg),\qquad v^{R}_t=\log\bigg(\frac{1}{n_t}\sum_{d\in t}\bar R_d^{\,2}\bigg),\qquad v_t=\log\bigg(\frac{1}{n_t}\sum_{d\in t}\sigma^2_{Q,d}\bigg).
\end{equation}
We refer to $y_t$ as the \emph{chaos} of month $t$, to $v^{R}_t$ as its realized market variance, and to $v_t$ as its implied variance.

\begin{definition}[Terms of the chaos accounting]\label{def:frag}
Let $y_t$ be the monthly index~\eqref{eq:monthly}, and let $y^{s}_t$, $v^{R}_t$ and $v_t$ be the monthly series~\eqref{eq:monthlyseries}. The fragmentation ratio, the wedge, the co-movement term and the form gap of month $t$ are
\begin{equation}\label{eq:terms}
f_t=y_t-v_t,\qquad w_t=v^{R}_t-v_t,\qquad c_t=y^{s}_t-v^{R}_t,\qquad g_t=y_t-y^{s}_t,
\end{equation}
respectively, and the chaos-implied correlation of month $t$ is
\begin{equation}\label{eq:rhohat}
\hat\rho_t=\frac{1}{1+e^{f_t}}.
\end{equation}
\end{definition}

The four series~\eqref{eq:terms} add up by construction, and Theorem~\ref{thm:dci}(ii) identifies the co-movement term. We thus obtain, for every month,
\begin{equation}\label{eq:accounting}
y_t=v_t+w_t+c_t+g_t,\qquad c_t=-\logit\tilde\rho_t,
\end{equation}
which we refer to as the \emph{chaos accounting}. Every term of the chaos accounting~\eqref{eq:accounting} is observed. The implied variance $v_t$ is the part of chaos which the VIX registers. The wedge $w_t$ is the log ratio of the realized variance of the equal-weighted market to the implied variance of the index. It moves for three reasons, namely the breadth of the turmoil, since the equal-weighted market of all assets can be more or less volatile than the capitalization-weighted index of the largest ones, the variance risk premium, which places implied above expected variance, and the surprise of the month, by which realized variance differs from the variance that was expected. The co-movement term $c_t$ is the log odds against co-movement, and it rises when stocks pull apart. The form gap $g_t$ is the log ratio of the logarithmic to the simple-return form of the monthly index, which by Theorem~\ref{thm:ustat}(v) reflects mainly the skewness of the cross-section and is large in months in which single stocks collapse. The fragmentation ratio collects the three terms which the VIX does not register, and the chaos-implied correlation~\eqref{eq:rhohat} satisfies
\begin{equation}\label{eq:rhohatlogit}
f_t=w_t+c_t+g_t,\qquad \logit\hat\rho_t=\logit\tilde\rho_t-w_t-g_t.
\end{equation}
The chaos-implied correlation is therefore the realized co-movement ratio on the logit scale, shifted by the wedge and by the form gap. It requires the index and the VIX alone, and it equals the realized co-movement ratio when the wedge and the form gap vanish.

The breadth component of the wedge arises because Theorem~\ref{thm:dci} weights all assets equally, whereas the VIX refers to an index with capitalization weights. The next proposition shows that the relation between dispersion and co-movement, its realized counterpart and the closed form of the index hold for arbitrary weights.

\begin{proposition}[Weighted identity]\label{prop:weighted}
Let $\bm\varpi\in\R^N$ be a vector of nonnegative weights with sum one, let $\vR$ be a vector of simple returns with conditional mean $\bm\mu$ and conditional covariance matrix $\mSigma$ given $\mathcal F$, and let
\begin{equation*}
R_\varpi=\bm\varpi^\top\vR,\qquad \CSV_\varpi=\sum_{i=1}^N\varpi_i\,(R_i-R_\varpi)^2,\qquad \bar\sigma^2_\varpi=\sum_{i=1}^N\varpi_i\,\Sigma_{ii},\qquad \bar\rho_\varpi=\frac{\bm\varpi^\top\mSigma\bm\varpi}{\bar\sigma^2_\varpi}.
\end{equation*}
Then, the following hold.
\begin{enumerate}[label=\normalfont(\roman*)]
\item The weighted co-movement ratio satisfies $0\le\bar\rho_\varpi\le1$, and
\begin{equation*}
\E\big[\CSV_\varpi\mid\mathcal F\big]=\bar\sigma^2_\varpi\,(1-\bar\rho_\varpi)+\sum_{i=1}^N\varpi_i\,\big(\mu_i-\bm\varpi^\top\bm\mu\big)^2.
\end{equation*}
\item For every realization of the returns,
\begin{equation*}
\sum_{i=1}^N\varpi_i\,R_i^2=R_\varpi^2+\CSV_\varpi,\qquad \CSV_\varpi=\sum_{i<j}\varpi_i\varpi_j\,(R_i-R_j)^2.
\end{equation*}
\item The weighted average of the elements of the pairwise comparison matrix is the ratio of the weighted arithmetic mean to the weighted harmonic mean of the gross returns, that is,
\begin{equation*}
\sum_{i,j=1}^N\varpi_i\varpi_j\,a_{ij}=\Big(\sum_{j=1}^N\varpi_j\,r_j\Big)\Big(\sum_{i=1}^N\frac{\varpi_i}{r_i}\Big).
\end{equation*}
\end{enumerate}
\end{proposition}

The proof is given in \ref{app:proofs}. By part~(ii), the realized identity of Theorem~\ref{thm:dci}(ii) holds with the weighted return $R_\varpi$ and the weighted dispersion $\CSV_\varpi$ in place of their equal-weighted counterparts. When $\bm\varpi$ contains the weights of the S\&P~500, the realized market variance of the chaos accounting~\eqref{eq:accounting} is the realized variance of the index itself, to which the VIX refers. The wedge then compares the realized and the implied variance of one portfolio, and its breadth component is absent. Part~(iii) is the weighted version of the closed form~\eqref{eq:fcix}. The empirical analysis uses equal weights, with which the index was defined~\cite{ataei2021theory}.

The chaos accounting~\eqref{eq:accounting} is linear, so that it also holds for changes. We use it to classify months of high chaos by the term that leads them.

\begin{definition}[Two faces]\label{def:faces}
For a monthly series $z_t$, let
\begin{equation}\label{eq:excess}
\Delta z_t=z_t-\frac{1}{12}\sum_{k=1}^{12}z_{t-k}
\end{equation}
be its excess over the mean of the preceding twelve months. A month $t$ is called synchronized if $\Delta v_t\ge\Delta f_t$, and fragmented otherwise. The leading term of a fragmented month is the largest of $\Delta w_t$, $\Delta c_t$ and $\Delta g_t$.
\end{definition}

By the chaos accounting~\eqref{eq:accounting}, the excess chaos of a month is the sum $\Delta v_t+\Delta w_t+\Delta c_t+\Delta g_t$. When the excess chaos is positive, implied variance accounts for at least half of it in a synchronized month, and the VIX registers the turmoil. In a fragmented month, the larger part of the excess chaos is not registered by the VIX, and the leading term states why. When the co-movement term leads, stocks pull apart, and the market index can be calm while the cross-section is in turmoil. We then speak of fragmentation in the narrow sense. When the wedge leads, the broad market moves more than index options had priced. When the form gap leads, the excess stems from the collapse of single stocks. We also refer to $\Delta v_t$ and $\Delta f_t$ as the systematic and the fragmentation channel of excess chaos, respectively.

\begin{example}\label{ex:onefactor-faces}
Let the returns of the $N$ assets be generated by a one-factor market. Given the loadings, the covariance matrix of the returns~\eqref{eq:onefactor} is $\sigma_F^2\,\vb\vb^\top+\sigma_\varepsilon^2\bm I$, with $\vb$ the vector of loadings, and with the mean and the variance of the loadings at their population values the quantities~\eqref{eq:comove} are
\begin{equation*}
\bar\sigma^2=\sigma_F^2\big(\bar b^2+s_b^2\big)+\sigma_\varepsilon^2,\qquad \sigma^2_{\mathrm{EW}}=\sigma_F^2\,\bar b^2+\frac{\sigma_\varepsilon^2}{N}.
\end{equation*}
By Theorem~\ref{thm:dci}(i), the expected dispersion is $\sigma_F^2s_b^2+\sigma_\varepsilon^2(1-1/N)$. At the parameter values of Example~\ref{ex:onefactor-index}, the co-movement ratio is $\bar\rho=0.197$, and the expected dispersion corresponds to a cross-sectional standard deviation of $0.020$. Let implied variance equal the variance of the equal-weighted portfolio, so that the wedge vanishes, and consider two shocks. In the first, the volatility $\sigma_F$ of the factor doubles. The logarithm of the market variance rises by $1.383$, the logarithm of the expected dispersion rises by $0.064$, and the co-movement ratio rises to $0.479$. In the second, the idiosyncratic volatility $\sigma_\varepsilon$ rises by one half. The logarithm of the expected dispersion rises by $0.799$, the logarithm of the market variance rises by $0.005$, and the co-movement ratio falls to $0.100$. The first shock is synchronized and the second is fragmented in the sense of Definition~\ref{def:faces}, and they are the two polar cases of which observed episodes are mixtures.
\end{example}

\section{Two Faces of Chaos}\label{sec:faces}

Section~\ref{sec:fcix} split chaos into four observed terms and defined the two faces. This section applies the chaos accounting~\eqref{eq:accounting} to large U.S.\ stocks from 1990 to 2025. It first describes the data and quantifies, on these data, the influence of single stocks and the sampling error of the monthly index. It then examines how well the chaos-implied correlation tracks co-movement, classifies the episodes of high chaos, describes the anatomy of the fragmented ones, and asks how the shares of the two faces depend on the way in which episodes are selected, on the form of the index and on the universe of stocks.

Returns were taken from the daily stock file of the Center for Research in Security Prices (CRSP), provided by Wharton Research Data Services, and the index was computed for the $9{,}067$ trading days from 2 January 1990 to 31 December 2025. The universe of day $d$ contains the $1{,}000$ largest U.S.\ common stocks listed on the NYSE, AMEX and Nasdaq with a price of at least five dollars, ranked by market capitalization on day $d-1$ among the stocks with a valid return on day $d$. Membership is therefore determined by information dated $d-1$ or earlier, except for the requirement of a valid return on day $d$, which omits the last-day returns of delisted stocks. The universe has no look-ahead in its composition and is not restricted to surviving firms. Returns are computed from close to close, gaps are never bridged, and the first return after a listing is excluded. This universe replaces the balanced panel of $811$ stocks with complete price histories used by Ataei et al.~\cite{ataei2021theory}, which Definition~\ref{def:fcix} makes unnecessary. Two alternative universes, namely the $500$ largest stocks and the point-in-time constituents of the S\&P~500, serve for robustness. The identities of Theorem~\ref{thm:ustat} were verified on every cross-section to machine precision.

Daily closing values of the VIX were obtained from Cboe. The sample runs from January 1990 to December 2025, which gives $T=432$ months. For each month, we computed the chaos $y_t$ of equation~\eqref{eq:monthly}, the series~\eqref{eq:monthlyseries} and the terms~\eqref{eq:terms} of the chaos accounting. Averaging the daily indexes, instead of computing the dispersion of monthly returns, measures the dispersion accumulated day by day, part of which the short-horizon reversal of relative returns~\cite{lo1990contrarian} removes from monthly returns. We also recorded two end-of-month values, namely the logarithm $v^{e}_t$ of the implied variance on the last trading day of month $t$ and the logarithm $y^{e}_t$ of the mean of the daily index over its last five trading days. A second, independent measure of co-movement is the Cboe one-month implied correlation index (COR1M), which is computed from the prices of index options and of options on individual stocks and is available for the $240$ months from 2006 to 2025.

\begin{figure}[!ht]
\centering
\includegraphics[width=\textwidth]{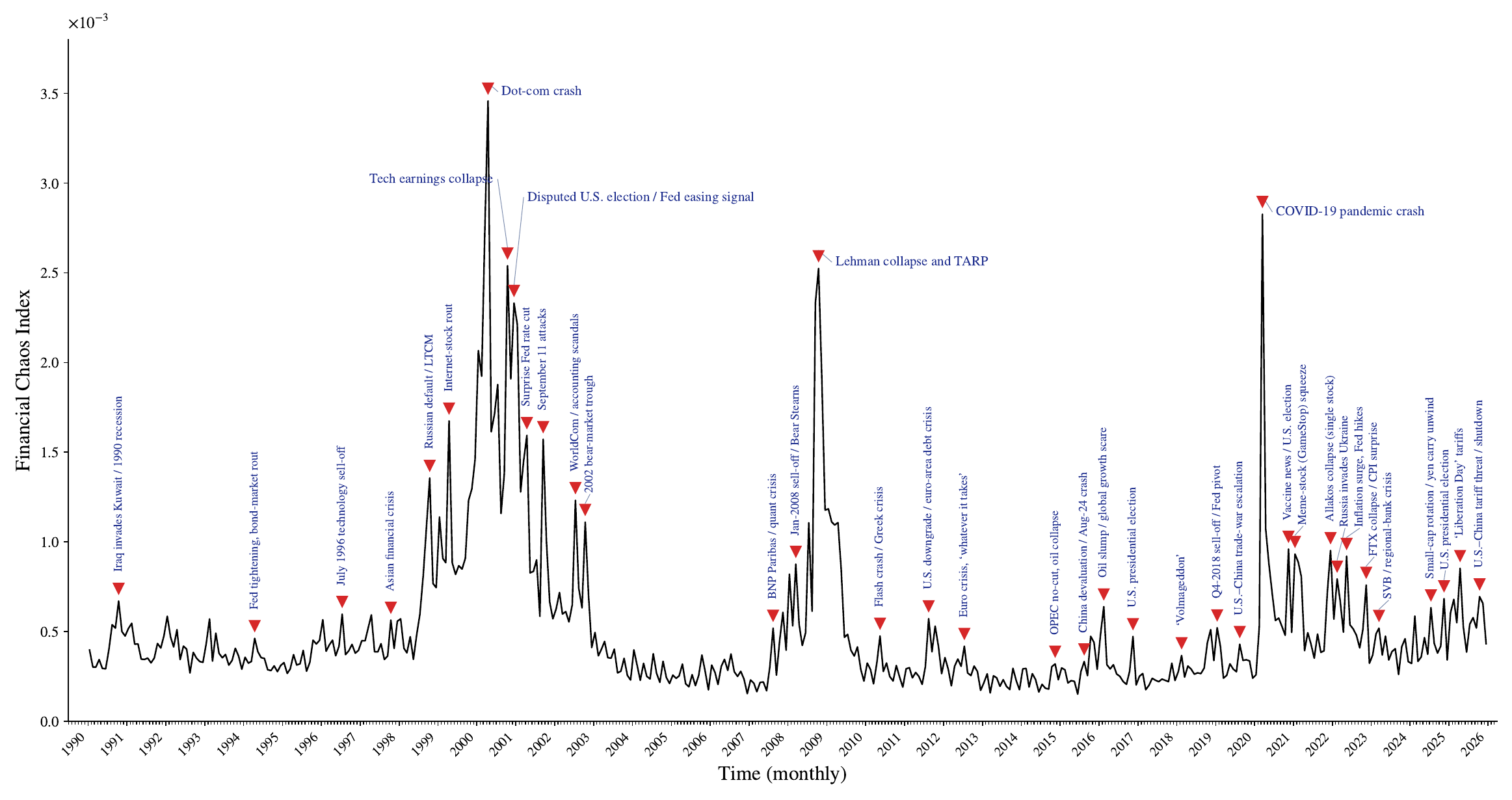}
\caption{Monthly financial chaos index in multiplicative form, 1990--2025. Triangles mark the 38 annotated events.}\label{fig:annot}
\end{figure}

Following Ataei et al.~\cite{ataei2021theory}, Figure~\ref{fig:annot} shows the monthly mean of the multiplicative index together with $38$ economic, financial, geopolitical and public-health events which coincide with its local peaks. Three features stand out. The largest monthly value occurs in April 2000, and not in 2008 or in 2020. The years from 1998 to 2002 form a long plateau of high chaos, whereas the episodes of 2008 and of 2020 are sharp and short. Finally, the level of the index after 2020 lies above that of the 2010s.

\begin{table}[!ht]
\centering\small
\caption{Monthly series of the chaos accounting, 1990--2025.}\label{tab:summary}
\begin{tabular}{@{}lccccc@{}}
\toprule
Series & Mean & Standard deviation & Minimum & Maximum & Autocorrelation\\
\midrule
Chaos $y_t$ & $-7.781$ & $0.568$ & $-8.795$ & $-5.673$ & $0.833$\\
Implied variance $v_t$ & $-8.909$ & $0.662$ & $-10.106$ & $-6.448$ & $0.876$\\
Wedge $w_t$ & $-0.486$ & $0.527$ & $-1.855$ & $1.027$ & $0.426$\\
Co-movement term $c_t$ & $1.604$ & $0.744$ & $-0.797$ & $3.504$ & $0.616$\\
Form gap $g_t$ & $0.010$ & $0.044$ & $-0.146$ & $0.404$ & $0.093$\\
Fragmentation ratio $f_t$ & $1.128$ & $0.496$ & $-0.323$ & $2.574$ & $0.732$\\
\bottomrule
\end{tabular}
\end{table}

Table~\ref{tab:summary} reports the mean, the standard deviation, the range and the first-order autocorrelation of chaos, of the four terms of the chaos accounting~\eqref{eq:accounting}, and of the fragmentation ratio, for the $432$ months. The mean of chaos, $-7.781$, corresponds to a daily cross-sectional standard deviation of returns of $e^{-7.781/2}=0.020$. The mean wedge is negative, so that the realized variance of the equal-weighted market is on average the fraction $e^{-0.486}=0.615$ of the implied variance of the index, which is consistent with a positive variance risk premium~\cite{bollerslev2009expected}. The mean co-movement term, $1.604$, corresponds to a realized co-movement ratio whose mean and median over the months are $0.191$ and $0.160$, respectively. By the relation~\eqref{eq:rhohatlogit}, the negative mean of the wedge places the chaos-implied correlation above the realized co-movement ratio, with a mean of $0.255$ and a median of $0.242$. The mean of COR1M over its $240$ months is $0.379$. The form gap is small on average, with a mean of $0.010$ and a standard deviation of $0.044$, but its maximum of $0.404$ is of the size of a chaos episode.

The form gap and the gap of Theorem~\ref{thm:ustat}(iv) are small on most days and large on a few, on which one stock dominates the index. The log ratio of $(N-1)\F_d/N$ to $\Fl_d$ has a median of $5.34\times10^{-4}$ and a 99th percentile of $0.018$, and it reaches $0.408$ on 22 December 2021, when one biotechnology stock lost $89.9\%$ of its value and accounted for $96\%$ of the logarithmic index of the day. On that day, the index was $5.49\times10^{-3}$ in its logarithmic form and $1.04\times10^{-3}$ in its simple-return form. A single stock accounts for more than half of the logarithmic index on $92$ of the $9{,}067$ days. At the monthly frequency, the form gap exceeds $0.1$ in $13$ months and $0.2$ in two, namely September 2008, with $0.288$, and December 2021, with $0.404$. The logarithmic and the simple-return form of the monthly index are otherwise close, with a correlation of $0.997$ between $y_t$ and $y^{s}_t$.

The sampling error of the monthly index follows from Theorem~\ref{thm:clt}. Since every term of the series~\eqref{eq:gap} is nonnegative, inverting its leading term day by day bounds the realized cross-sectional kurtosis of the log returns from above. The median of this bound over the days is $13.3$, and its 90th percentile is $54.5$. By Theorem~\ref{thm:clt}(iv), the realized kurtosis in turn overstates the effective kurtosis. Using the daily bounds in the variance formula~\eqref{eq:measerr} with $N=10^3$, and the daily indexes in place of the dispersion variances, we find that the standard deviation of the sampling error of $y_t$ is at most $0.041$ in the median month and at most $0.076$ at the 90th percentile of the months. The bound exceeds $0.1$ in $5.6\%$ of the months, almost all of which are months in which a single stock collapsed, and it reaches $0.336$ in December 2021. Two caveats apply. The bound rests on the independence of the idiosyncratic shocks across assets, and a clustering of shocks within industries would raise the effective kurtosis. Moreover, by Remark~\ref{rem:fourth}, the bound loses its meaning on days on which the fourth moment of the cross-section is not finite, of which the days dominated by one stock are the visible instances. In the remaining months, the sampling error is small relative to the standard deviation $0.568$ of chaos.

\begin{table}[!ht]
\centering\small
\caption{Chaos-implied correlation against two measures of co-movement.}\label{tab:valid}
\begin{tabular}{@{}lccccccc@{}}
\toprule
 & & \multicolumn{3}{c}{$R^2$ in levels} & \multicolumn{3}{c}{$R^2$ in monthly changes}\\
\cmidrule(lr){3-5}\cmidrule(lr){6-8}
Regressand & Months & $v_t$ & $\hat\rho_t$ & both & $v_t$ & $\hat\rho_t$ & both\\
\midrule
COR1M (implied) & 240 & 0.460 & 0.690 & 0.713 (7.50) & 0.545 & 0.296 & 0.567 (2.75)\\
Realized co-movement ratio $\tilde\rho_t$ & 432 & 0.359 & 0.514 & 0.562 (7.84) & 0.468 & 0.280 & 0.498 (2.95)\\
\bottomrule
\end{tabular}
\end{table}

\begin{figure}[!ht]
\centering
\includegraphics[width=\textwidth]{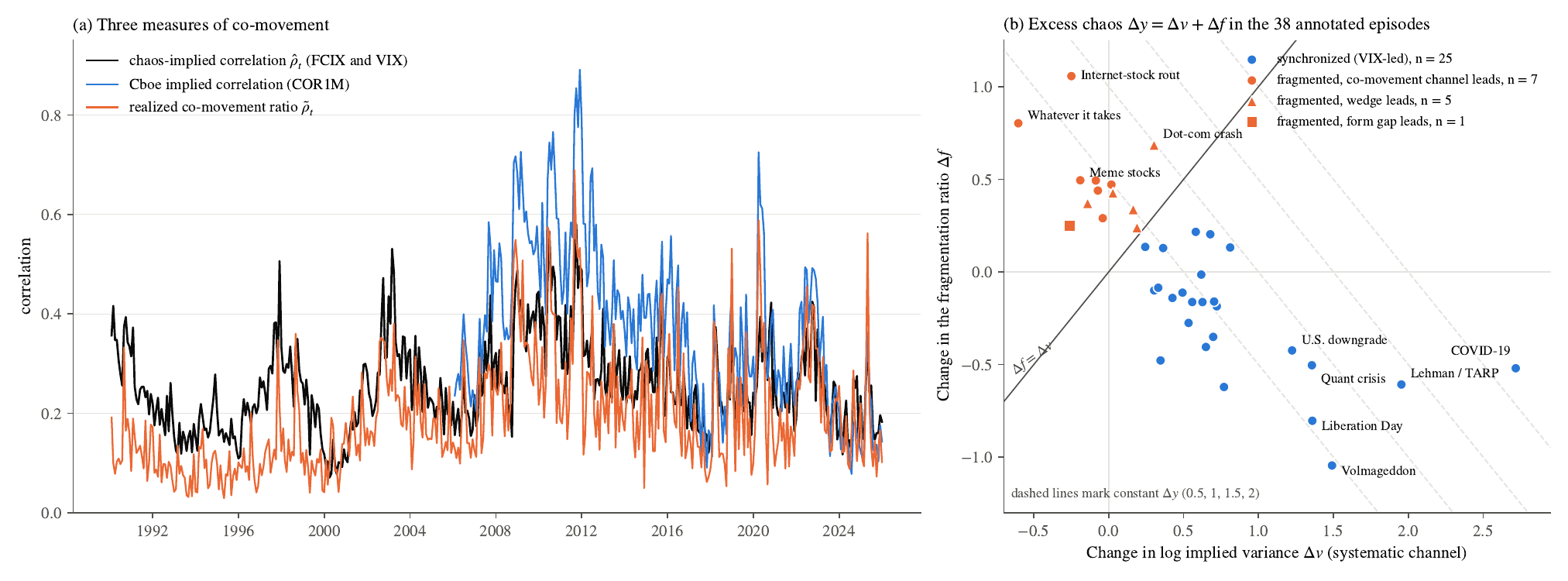}
\caption{Two faces of chaos. (a) Chaos-implied correlation (black), COR1M (blue) and realized co-movement ratio (orange), 1990--2025. (b) Excess implied variance and excess fragmentation ratio of the 38 annotated episodes, with synchronized episodes in blue and fragmented episodes in orange, the marker of a fragmented episode showing its leading term.}\label{fig:faces}
\end{figure}

The chaos-implied correlation~\eqref{eq:rhohat} uses the index and the VIX alone. Table~\ref{tab:valid} compares it with two measures of co-movement which use other information, namely COR1M, which is implied by option prices, and the realized co-movement ratio~\eqref{eq:rhotilde}. Each measure is regressed on implied variance $v_t$ alone, on the chaos-implied correlation $\hat\rho_t$ alone, and on both, in levels and in monthly changes, and the table reports the coefficients of determination and, in parentheses, the $t$-statistic of $\hat\rho_t$ in the joint regression, with Newey--West standard errors~\cite{newey1987simple} and $12$ lags. In levels, the chaos-implied correlation has a correlation of $0.831$ with COR1M and of $0.717$ with the realized co-movement ratio, and it explains more of either measure than implied variance does. The three measures are shown in Figure~\ref{fig:faces}(a). In monthly changes, the order is reversed. Implied variance alone explains $0.545$ of the variance of the changes of COR1M, and adding the chaos-implied correlation raises the share to $0.567$. The chaos-implied correlation thus tracks the level and the slow movements of co-movement, whereas its month-to-month changes add little to those of the VIX. By the relation~\eqref{eq:rhohatlogit}, the difference between the chaos-implied correlation and the realized co-movement ratio is the wedge, apart from the small form gap, and the wedge has a standard deviation of $0.527$, which is of the same order as that of the co-movement term (see Table~\ref{tab:summary}). The wedge is therefore a first-order term of the accounting, and the episodes are classified with all four terms in view.

We consider the $38$ annotated events of Figure~\ref{fig:annot}, and we measure each by the excesses~\eqref{eq:excess} of the month of the event. For the first event, in October 1990, the mean is taken over the nine preceding months of the sample. Table~\ref{tab:episodes} lists, for every event, the excess chaos $\Delta y_t$ and its four terms, the face according to Definition~\ref{def:faces}, with S for synchronized and F for fragmented, and the leading term of each fragmented episode. Panel~A of Table~\ref{tab:faces} reports the means by face, together with the number of episodes in which the co-movement term rises. Of the $38$ episodes, $25$ are synchronized and $13$ are fragmented, a share of $13/38=0.342$ with an exact $95\%$ confidence interval from $0.196$ to $0.514$.

\begin{table}[!ht]
\centering\footnotesize
\caption{Chaos accounting of the 38 annotated episodes.}\label{tab:episodes}
\begin{tabular}{@{}llcccccll@{}}
\toprule
Month & Event & $\Delta y$ & $\Delta v$ & $\Delta w$ & $\Delta c$ & $\Delta g$ & Face & Leading term\\
\midrule
1990-10 & Iraq invades Kuwait / 1990 recession & $0.604$ & $0.618$ & $0.420$ & $-0.438$ & $0.003$ & S & \\
1994-04 & Fed tightening, bond-market rout & $0.258$ & $0.533$ & $0.805$ & $-1.077$ & $-0.004$ & S & \\
1996-07 & July 1996 technology sell-off & $0.381$ & $0.493$ & $0.874$ & $-1.015$ & $0.029$ & S & \\
1997-10 & Asian financial crisis & $0.285$ & $0.425$ & $1.449$ & $-1.640$ & $0.051$ & S & \\
1998-10 & Russian default / LTCM & $0.943$ & $0.811$ & $0.442$ & $-0.295$ & $-0.015$ & S & \\
1999-04 & Internet-stock rout & $0.808$ & $-0.249$ & $-0.289$ & $1.355$ & $-0.008$ & F & co-movement\\
2000-04 & Dot-com crash & $0.988$ & $0.303$ & $1.692$ & $-1.037$ & $0.029$ & F & wedge\\
2000-10 & Tech earnings collapse & $0.379$ & $0.244$ & $0.606$ & $-0.546$ & $0.075$ & S & \\
2000-12 & Disputed U.S.\ election / Fed easing signal & $0.204$ & $0.304$ & $0.655$ & $-0.757$ & $0.003$ & S & \\
2001-04 & Surprise Fed rate cut & $-0.132$ & $0.346$ & $0.368$ & $-0.799$ & $-0.046$ & S & \\
2001-09 & September 11 attacks & $0.149$ & $0.770$ & $0.178$ & $-0.880$ & $0.081$ & S & \\
2002-07 & WorldCom / accounting scandals & $0.536$ & $0.721$ & $0.760$ & $-0.988$ & $0.043$ & S & \\
2002-10 & 2002 bear-market trough & $0.463$ & $0.626$ & $0.538$ & $-0.719$ & $0.018$ & S & \\
2007-08 & BNP Paribas / quant crisis & $0.853$ & $1.356$ & $0.208$ & $-0.734$ & $0.022$ & S & \\
2008-03 & Jan-2008 sell-off / Bear Stearns & $0.882$ & $0.678$ & $0.539$ & $-0.525$ & $0.189$ & S & \\
2008-10 & Lehman collapse and TARP & $1.346$ & $1.954$ & $0.631$ & $-1.211$ & $-0.027$ & S & \\
2010-05 & Flash crash / Greek crisis & $0.244$ & $0.649$ & $0.613$ & $-1.217$ & $0.199$ & S & \\
2011-08 & U.S.\ downgrade / euro-area debt crisis & $0.801$ & $1.224$ & $1.328$ & $-1.858$ & $0.106$ & S & \\
2012-07 & Euro crisis, `whatever it takes' & $0.199$ & $-0.604$ & $-0.024$ & $0.827$ & $0.001$ & F & co-movement\\
2014-11 & OPEC no-cut, oil collapse & $0.367$ & $-0.072$ & $-1.170$ & $1.565$ & $0.045$ & F & co-movement\\
2015-08 & China devaluation / Aug-24 crash & $0.347$ & $0.698$ & $0.821$ & $-1.194$ & $0.023$ & S & \\
2016-02 & Oil slump / global growth scare & $0.798$ & $0.581$ & $0.293$ & $-0.125$ & $0.049$ & S & \\
2016-11 & U.S.\ presidential election & $0.408$ & $-0.086$ & $-0.105$ & $0.629$ & $-0.030$ & F & co-movement\\
2018-02 & `Volmageddon' & $0.446$ & $1.491$ & $0.539$ & $-1.605$ & $0.021$ & S & \\
2019-01 & Q4-2018 sell-off / Fed pivot & $0.493$ & $0.363$ & $0.371$ & $-0.237$ & $-0.004$ & S & \\
2019-08 & U.S.--China trade-war escalation & $0.246$ & $0.331$ & $0.789$ & $-0.921$ & $0.047$ & S & \\
2020-03 & COVID-19 pandemic crash & $2.198$ & $2.717$ & $1.318$ & $-1.886$ & $0.048$ & S & \\
2020-11 & Vaccine news / U.S.\ election & $0.490$ & $0.018$ & $0.003$ & $0.489$ & $-0.020$ & F & co-movement\\
2021-01 & Meme-stock (GameStop) squeeze & $0.305$ & $-0.190$ & $-0.240$ & $0.819$ & $-0.084$ & F & co-movement\\
2021-12 & Allakos collapse (single stock) & $0.429$ & $0.189$ & $0.482$ & $-0.656$ & $0.414$ & F & wedge\\
2022-02 & Russia invades Ukraine & $0.395$ & $0.558$ & $0.343$ & $-0.538$ & $0.033$ & S & \\
2022-05 & Inflation surge, Fed hikes & $0.544$ & $0.704$ & $0.727$ & $-0.940$ & $0.053$ & S & \\
2022-11 & FTX collapse / CPI surprise & $0.230$ & $-0.141$ & $0.518$ & $-0.145$ & $-0.002$ & F & wedge\\
2023-03 & SVB / regional-bank crisis & $-0.011$ & $-0.261$ & $0.086$ & $0.050$ & $0.114$ & F & form gap\\
2024-07 & Small-cap rotation / yen carry unwind & $0.455$ & $0.028$ & $0.291$ & $-0.027$ & $0.163$ & F & wedge\\
2024-11 & U.S.\ presidential election & $0.502$ & $0.164$ & $0.355$ & $0.014$ & $-0.030$ & F & wedge\\
2025-04 & `Liberation Day' tariffs & $0.556$ & $1.359$ & $1.064$ & $-1.976$ & $0.108$ & S & \\
2025-10 & U.S.--China tariff threat / shutdown & $0.250$ & $-0.040$ & $0.052$ & $0.217$ & $0.021$ & F & co-movement\\
\bottomrule
\end{tabular}
\end{table}

\begin{table}[!ht]
\centering\small
\caption{Mean excess chaos and its terms by face.}\label{tab:faces}
\begin{tabular}{@{}lcccccccc@{}}
\toprule
 & Episodes & $\Delta y$ & $\Delta v$ & $\Delta f$ & $\Delta w$ & $\Delta c$ & $\Delta g$ & $\Delta c>0$\\
\midrule
\multicolumn{9}{@{}l}{\emph{Panel A. The 38 annotated episodes}}\\
\quad Synchronized (VIX-led) & 25 & $0.569$ & $0.822$ & $-0.254$ & $0.667$ & $-0.965$ & $0.044$ & 0\\
\quad Fragmented (dispersion-led) & 13 & $0.417$ & $-0.072$ & $0.489$ & $0.127$ & $0.315$ & $0.047$ & 9\\
\midrule
\multicolumn{9}{@{}l}{\emph{Panel B. The 35 rule-based episodes}}\\
\quad Synchronized (VIX-led) & 18 & $0.664$ & $0.827$ & $-0.163$ & $0.645$ & $-0.847$ & $0.039$ & 0\\
\quad Fragmented (dispersion-led) & 17 & $0.546$ & $0.042$ & $0.504$ & $0.265$ & $0.192$ & $0.047$ & 10\\
\bottomrule
\end{tabular}
\end{table}

The two faces differ in every term of the accounting. In the synchronized episodes, implied variance rises by $0.822$ on average, and the realized variance of the market rises by more, so that the wedge rises by $0.667$. Co-movement rises in each of the $25$ episodes, and the co-movement term falls by $0.965$ on average. Dispersion rises nevertheless, by $0.569$, because the variance of the market rises by more than the odds against co-movement fall. In the fragmented episodes, implied variance does not rise on average. Its excess is negative in $8$ of the $13$ episodes and never exceeds $0.303$. The excess chaos of $0.417$ is carried by the co-movement term, which rises by $0.315$ on average and in $9$ of the $13$ episodes, and by the wedge, which rises by $0.127$. Figure~\ref{fig:faces}(b) shows the separation of the two groups. The two faces are of the same order of magnitude. The mean excess chaos is $0.569$ in the synchronized and $0.417$ in the fragmented episodes, with medians of $0.463$ and $0.408$, respectively, and a Welch test does not reject the equality of the means, with a $p$-value of $0.201$. Newspapers register the fragmented episodes as little as the VIX does. Relative to the preceding twelve months, the logarithm of the newspaper-based tracker of equity market volatility of Baker et al.~\cite{baker2026policy} rises by $0.529$ in the synchronized episodes and falls by $0.011$ in the fragmented ones.

The leading terms show that the fragmented episodes are of three kinds. In seven of them, the co-movement term leads, and stocks pull apart while the market as a whole is calm. The internet-stock rout of April 1999 is the clearest case, with an excess chaos of $0.808$ which consists of a co-movement term of $1.355$ and of negative excesses of implied variance and of the wedge. The others are July 2012, November 2014, November 2016, November 2020, the meme-stock squeeze of January 2021, and October 2025. In five episodes, the wedge leads. In the dot-com crash of April 2000, the month of the highest chaos of the sample, the co-movement term fell by $1.037$, so that stocks moved together, while the wedge rose by $1.692$. The broad market was in turmoil to an extent which the options on the S\&P~500 did not price. Measured against the equal-weighted return of the constituents of the S\&P~500, $0.907$ of this wedge is the excess of the realized variance of the $1{,}000$ stocks over that of the constituents, and $0.786$ is the excess of the realized variance of the constituents over implied variance. The other wedge-led episodes are December 2021, November 2022, July 2024 and November 2024. In the remaining episode, the regional-bank crisis of March 2023, the form gap leads, and there is no excess chaos to explain.

December 2021 requires a separate comment. Its excess chaos of $0.429$ consists of an excess implied variance of $0.189$, a wedge of $0.482$, a co-movement term of $-0.656$ and a form gap of $0.414$. The form gap is the contribution of the biotechnology stock which collapsed on 22 December, and without it the month is not an episode. The excess of the simple-return form of the monthly index is $0.015$, against $0.429$ for the logarithmic form and $0.582$ for the multiplicative form. The peak of the index in that month reflects the sensitivity of the logarithmic and of the multiplicative form to one extreme return, in line with Theorem~\ref{thm:ustat}(v), and not a state of the market.

The $38$ events were chosen by inspection of the index, and the choice may favor either face. We therefore repeat the classification for episodes selected by a rule, and for all months. The rule selects the months whose excess chaos exceeds $0.3$ and is the largest within two months on either side, where the threshold is more than seven times the median bound $0.041$ on the standard deviation of the sampling error. It selects $35$ episodes, of which $17$ are fragmented (Panel~B of Table~\ref{tab:faces}). Their accounting resembles that of the annotated episodes. Implied variance rises by $0.042$ on average in the fragmented and by $0.827$ in the synchronized episodes, and the co-movement term rises in $10$ of the $17$ fragmented episodes and in none of the synchronized ones. The co-movement term leads in $9$ and the wedge in $8$ of the fragmented episodes. The wedge-led group includes September 2008, the month of the failure of Lehman Brothers, in which the realized variance of the market outran the monthly mean of implied variance. A fragmented month in the sense of Definition~\ref{def:faces} is therefore a month whose turmoil the VIX under-registers, and it is a month of falling co-movement in about half of the cases, namely in $7$ of $13$ and in $9$ of $17$ episodes by the leading term.

Among all $420$ months for which the excess~\eqref{eq:excess} is defined, the excess chaos exceeds $0.3$ in $72$, and $29$ of these, a share of $0.403$, are fragmented. In the regression of excess chaos on excess implied variance over the $420$ months, the slope is $0.492$ and the coefficient of determination is $0.421$. Implied variance thus accounts for less than half of the variance of excess chaos. The variance of the excess fragmentation ratio, $0.142$, is much smaller than the sum of the variances of its terms, which are $0.213$ for the wedge, $0.355$ for the co-movement term and $0.002$ for the form gap, because the wedge and the co-movement term have a correlation of $-0.773$. Part of this negative correlation is mechanical, since the realized market variance $v^{R}_t$, which is the logarithm of a mean of about $21$ squared daily returns, enters the two terms with opposite signs.

The selection matters for one conclusion. Among the annotated episodes, $2$ of the $18$ before 2012 and $11$ of the $20$ since 2012 are fragmented, which suggests that fragmented chaos has become the dominant form. The suggestion does not survive the selection by rule. Of the rule-based episodes, $7$ of the $19$ before 2012 and $10$ of the $16$ since 2012 are fragmented, and Fisher's exact test of equal shares has a $p$-value of $0.181$. Of the months of high chaos, $15$ of $42$ are fragmented from 1991 to 2011 and $14$ of $30$ from 2012 to 2025, and the shares in the two halves of the sample are $12/29=0.414$ for 1991--2007 and $17/43=0.395$ for 2008--2025. Fragmented chaos is thus present throughout the sample. What has changed is the association between chaos and the VIX, since the correlation between excess chaos and excess implied variance is $0.463$ before 2008 and $0.745$ thereafter.

\begin{table}[!ht]
\centering\small
\caption{Two faces by form of the index and universe.}\label{tab:robust}
\begin{tabular}{@{}lccccc@{}}
\toprule
 & \multicolumn{2}{c}{Fragmented episodes} & Fragmented & \multicolumn{2}{c}{Correlation of $\hat\rho_t$ with}\\
\cmidrule(lr){2-3}\cmidrule(lr){5-6}
Form and universe & annotated & rule-based & months & COR1M & $\tilde\rho_t$\\
\midrule
Top 1,000, logarithmic form (baseline) & 13/38 & 17/35 & 0.403 & 0.831 & 0.717\\
Top 1,000, multiplicative form & 13/38 & 17/35 & 0.411 & 0.829 & 0.714\\
Top 1,000, simple-return form & 12/38 & 12/32 & 0.373 & 0.836 & 0.734\\
Top 500, logarithmic form & 12/38 & 19/38 & 0.400 & 0.828 & 0.726\\
S\&P 500 constituents, logarithmic form & 12/38 & 16/35 & 0.338 & 0.660 & 0.727\\
\bottomrule
\end{tabular}
\end{table}

Table~\ref{tab:robust} repeats the classification for the multiplicative and the simple-return form of the index and for two narrower universes. It reports the number of fragmented episodes among the annotated and among the rule-based episodes, the share of fragmented months among the months of high chaos, and the correlations of the chaos-implied correlation with COR1M and with the realized co-movement ratio. The shares of the two faces change little. With the simple-return form, which removes the form gap, $12$ of the $38$ annotated and $12$ of the $32$ rule-based episodes are fragmented. With the $500$ largest stocks and with the constituents of the S\&P~500, the share of fragmented months is $0.400$ and $0.338$, respectively. The correlation of the chaos-implied correlation with the realized co-movement ratio lies between $0.714$ and $0.734$ in all variants, and its correlation with COR1M is $0.660$ for the constituents of the S\&P~500, against $0.831$ for the baseline.

\section{News and the Two Faces}\label{sec:news}

Section~\ref{sec:faces} found that newspapers register fragmented episodes as little as the VIX does. This section examines the relation between news and the terms of the chaos accounting month by month. It introduces the news data, shows why the intensity and the composition of news must be entered separately, and then asks which news accompanies each term in the same month and whether the news of the previous month predicts chaos. The first question is descriptive. Newspapers write about the moves of the market, so that the associations of news with the chaos of the same month have no causal interpretation.

We use the monthly Equity Market Volatility (EMV) trackers of Baker et al.~\cite{baker2026policy}. The overall tracker $E_t$ is a scaled count of the newspaper articles which contain terms on the economy, on the stock market and on volatility, and each of $40$ category trackers equals
\begin{equation}\label{eq:emv}
E_{j,t}=E_t\,s_{j,t},\qquad j=1,\dots,40,
\end{equation}
where $s_{j,t}\in[0,1]$ is the share of these articles which also contain the terms of category $j$. Every category tracker is thus the product of a common intensity and of a share which describes the composition of the news. In the source file, the ratio of $E_{j,t}$ to $E_t$ lies between zero and one for every category and month of 1990--2025, with the exception of the tracker of infectious disease in September 2021, whose ratio of $1.32$ we set to one. Four further trackers aggregate some of the $40$ categories and are not used. The terms of the EMV trackers were chosen so that the overall tracker follows the VIX~\cite{baker2026policy}. An association of the trackers with implied variance therefore recovers their construction in part, whereas the index is built from stock prices alone.

Entering the logarithms of the category trackers in a regression is tempting, but the product structure~\eqref{eq:emv} confounds intensity with composition, as the next proposition shows.

\begin{proposition}[Aliasing]\label{prop:alias}
Let $\bm{\mathcal E}_t=\iota_t\one+\bm\varsigma_t$ be the vector of the logarithms of $K$ category trackers, where $\iota_t=\log E_t$ and $\bm\varsigma_t$ is the vector of the logarithms of the shares, and suppose that a series $z_t$ satisfies
\begin{equation*}
z_t=\beta_0+\beta_\iota\,\iota_t+\bm\eta^\top\bm\varsigma_t+u_t,
\end{equation*}
where $u_t$ is uncorrelated with $\iota_t$ and $\bm\varsigma_t$. Then, the following hold.
\begin{enumerate}[label=\normalfont(\roman*)]
\item The regression of $z_t$ on $\bm{\mathcal E}_t$ is correctly specified for all values of $\iota_t$ and $\bm\varsigma_t$ if and only if
\begin{equation*}
\beta_\iota=\one^\top\bm\eta.
\end{equation*}
\item If $\bm\eta=\bm 0$, and if $\iota_t$ has variance $\sigma_\iota^2$ and is uncorrelated with $\bm\varsigma_t$, whose covariance matrix $\mSigma_\varsigma$ is positive definite, then the population coefficients of the regression of $z_t$ on $\bm{\mathcal E}_t$ and their sum are
\begin{equation}\label{eq:alias}
\bm\beta^{\mathrm{raw}}=\frac{\beta_\iota\,\sigma_\iota^2}{1+\varrho}\,\mSigma_\varsigma^{-1}\one,\qquad \one^\top\bm\beta^{\mathrm{raw}}=\frac{\varrho}{1+\varrho}\,\beta_\iota,
\end{equation}
respectively, where $\varrho=\sigma_\iota^2\,\one^\top\mSigma_\varsigma^{-1}\one$.
\end{enumerate}
\end{proposition}

The proof is given in \ref{app:proofs}. Part~(i) shows that a regression on the category trackers imposes a restriction between the effects of intensity and of composition for which there is no reason. Part~(ii) shows what happens when only the intensity matters. By the coefficients~\eqref{eq:alias}, every category then receives a nonzero coefficient which is determined by the covariance matrix of the irrelevant shares, the total effect is attenuated, and individual coefficients can have the sign opposite to that of $\beta_\iota$. We therefore enter the intensity and the composition of the news separately.

\begin{definition}[News measures]\label{def:news}
The intensity of the news of month $t$ is $\iota_t=\log E_t$. The $40$ categories are partitioned into $K=13$ themes, and the mention rate of theme $k$ is the sum $M_{k,t}$ of the shares $s_{j,t}$ of its categories. The theme regressors $x_{k,t}$ are the standardized values of $\log(M_{k,t}+M^{\min}_k/2)$, where $M^{\min}_k$ is the smallest positive value of $M_{k,t}$ in the sample, and the concentration of the news is
\begin{equation}\label{eq:conc}
C_t=1-\frac{H_t}{\log K},\qquad H_t=-\sum_{k=1}^Kp_{k,t}\log p_{k,t},\qquad p_{k,t}=\frac{M_{k,t}}{\sum_{h=1}^KM_{h,t}}.
\end{equation}
\end{definition}

The intensity measures how much news on volatility there is, the theme regressors measure what the news is about, and the concentration measures whether a few themes dominate it. The concentration~\eqref{eq:conc} lies between zero, for a uniform composition, and one, when a single theme accounts for all mentions, and it does not depend on the intensity. Table~\ref{tab:themes} lists the themes with the category trackers which they contain and the mean of the mention rate over 1990--2025.

\begin{table}[!ht]
\centering\footnotesize
\caption{News themes and their category trackers.}\label{tab:themes}
\begin{tabular}{@{}lL{0.68\textwidth}c@{}}
\toprule
Theme & Category trackers & Mean rate\\
\midrule
Real activity & Broad quantity indicators, labor markets, consumer spending and sentiment, business investment and sentiment & $0.626$\\
Inflation \& rates & Inflation, interest rates & $0.594$\\
Housing \& credit & Real-estate markets, government-sponsored enterprises, housing and land management & $0.344$\\
Financial stress & Financial crises, other financial indicators, exchange rates, financial regulation & $0.276$\\
Monetary policy & Monetary policy & $0.303$\\
Fiscal policy & Taxes, government spending, deficits and debt, entitlement and welfare programs & $0.431$\\
Trade & Trade as a macroeconomic topic, trade policy & $0.062$\\
National security & National security policy & $0.145$\\
Elections \& governance & Elections and political governance, immigration & $0.042$\\
Commodities \& energy & Commodity markets, petroleum markets, energy and environmental regulation, agricultural policy & $0.661$\\
Infectious disease & Infectious disease & $0.102$\\
Health sector & Healthcare matters, healthcare policy, food and drug policy & $0.130$\\
Firm-level \& legal & Litigation matters, competition matters, intellectual property matters, labor disputes, competition policy, intellectual property policy, lawsuit and tort reform and Supreme Court decisions, labor regulations, other regulation, transportation, infrastructure and public utilities & $0.240$\\
\bottomrule
\end{tabular}
\end{table}

\begin{table}[!ht]
\centering\small
\caption{News and the terms of the chaos accounting.}\label{tab:news}
\begin{tabular}{@{}lccccc@{}}
\toprule
 & $y_t$ & $v_t$ & $f_t$ & $w_t$ & $c_t$\\
\midrule
\multicolumn{6}{@{}l}{\emph{Panel A. News of month $t$}}\\
Intensity $\iota_t$ & $0.151^{***}$ & $0.200^{***}$ & $-0.049^{**}$ & $0.198^{***}$ & $-0.261^{***}$\\
 & $(0.026)$ & $(0.020)$ & $(0.023)$ & $(0.031)$ & $(0.039)$\\
Concentration $C_t$ & $0.006$ & $0.051^{***}$ & $-0.045^{***}$ & $0.081^{***}$ & $-0.128^{***}$\\
 & $(0.014)$ & $(0.011)$ & $(0.014)$ & $(0.014)$ & $(0.020)$\\
$R^2$ & 0.849 & 0.897 & 0.688 & 0.428 & 0.638\\
Thirteen themes, Wald statistic & 26.0 & 34.8 & 20.9 & 56.6 & 66.8\\
\quad wild-bootstrap $p$-value & 0.384 & 0.201 & 0.581 & 0.015 & 0.004\\
\midrule
\multicolumn{6}{@{}l}{\emph{Panel B. News of month $t-1$}}\\
Intensity $\iota_{t-1}$ & $0.024$ & $0.036^{*}$ & $-0.012$ & $-0.018$ & $0.008$\\
 & $(0.022)$ & $(0.022)$ & $(0.025)$ & $(0.034)$ & $(0.046)$\\
Concentration $C_{t-1}$ & $-0.006$ & $0.015$ & $-0.021$ & $0.017$ & $-0.040^{*}$\\
 & $(0.017)$ & $(0.013)$ & $(0.015)$ & $(0.019)$ & $(0.022)$\\
Thirteen themes, Wald statistic & 33.5 & 12.2 & 16.8 & 21.0 & 27.1\\
\quad wild-bootstrap $p$-value & 0.174 & 0.899 & 0.742 & 0.522 & 0.312\\
\bottomrule
\end{tabular}
\end{table}

Table~\ref{tab:news} reports regressions of chaos, of implied variance, of the fragmentation ratio, of the wedge and of the co-movement term on the standardized intensity and concentration of the news of the same month (Panel~A) and of the previous month (Panel~B). Every regression contains a constant and the same controls, all dated $t-1$, namely the last value and the means of the last $3$ and of the last $12$ values of each of $y_t$, $v_t$ and $v^{R}_t$, and the end-of-month values $v^{e}_{t-1}$ and $y^{e}_{t-1}$. The controls treat the terms symmetrically, so that the coefficients inherit the chaos accounting~\eqref{eq:accounting}. The coefficient for $y_t$ is the sum of the coefficients for $v_t$ and $f_t$, and the coefficient for $f_t$ is the sum of the coefficients for $w_t$ and $c_t$ and of a coefficient for the form gap, which is not shown and does not exceed $0.014$ in absolute value. The sample covers the $419$ months from February 1991 to December 2025. Newey--West standard errors with $12$ lags are in parentheses, and one, two and three asterisks mark significance at the $10\%$, $5\%$ and $1\%$ levels. The last two rows of each panel add the $13$ theme regressors and report the Wald statistic of their joint significance together with its $p$-value from a wild bootstrap.

Months of intense news have more chaos, and the chaos is tilted toward the VIX. A rise of the intensity by one standard deviation comes with a rise of implied variance by $0.200$ and of chaos by $0.151$, so that the fragmentation ratio falls by $0.049$. The split of the fragmentation ratio shows what lies behind this small number. The wedge rises by $0.198$, and the co-movement term falls by $0.261$. In months of intense news, the realized variance of the market rises by twice as much as implied variance, and co-movement rises. The concentration of news has a sharper signature. It is unrelated to the size of chaos, with a coefficient of $0.006$ and a standard error of $0.014$, but it is related to every term. Months of concentrated news have a higher implied variance, by $0.051$, a higher wedge, by $0.081$, and a lower co-movement term, by $0.128$, and the three effects cancel in chaos. When one story dominates the news, the market moves more as a whole and stocks move more together, with no more dispersion. The association of concentration with the co-movement term is the direct evidence that concentrated news comes with co-movement, and it does not rest on the VIX. It is present in both halves of the sample, with a coefficient of $-0.086$ and a $t$-statistic of $-2.05$ for 1991--2007, and a coefficient of $-0.222$ and a $t$-statistic of $-5.67$ for 2008--2025. The association with implied variance, in contrast, is confined to the second half, with coefficients of $0.008$ and $0.110$, respectively. The association of concentration with the co-movement term is also robust to the form of the index and to the universe of stocks. For the multiplicative form, the simple-return form, the $500$ largest stocks and the constituents of the S\&P~500 of Table~\ref{tab:robust}, the coefficient is $-0.128$, $-0.128$, $-0.131$ and $-0.118$, with $t$-statistics of $-6.40$, $-6.46$, $-6.20$ and $-5.77$, respectively. The coefficient for the fragmentation ratio, which is $-0.045$ for the baseline, weakens to $-0.028$ and $-0.019$ for the two narrower universes, with $t$-statistics of $-1.75$ and $-1.08$.

Inference on the $13$ themes jointly requires care. A Wald test with a Newey--West covariance matrix of $13$ coefficients, estimated from $419$ observations with $12$ lags, rejects far too often when it is referred to the chi-squared distribution, in line with the general findings on such tests~\cite{kiefer2005new,lazarus2018har}. In $4{,}000$ simulated samples of the configuration of Table~\ref{tab:news}, with $419$ observations, a constant, the own lag and its two means, and $13$ persistent and cross-correlated regressors with no effect, the test rejected at the nominal $5\%$ level in $0.35$ of the samples. We therefore refer the Wald statistics to a wild bootstrap with a fixed design~\cite{goncalves2004bootstrapping}, in which the residuals of the regression without the themes are multiplied by independent signs~\cite{davidson2008wild} in $1{,}999$ replications. By this test, the themes of the month are jointly related to the wedge and to the co-movement term, with $p$-values of $0.015$ and $0.004$, and not to chaos, to implied variance or to the fragmentation ratio.

\begin{figure}[!ht]
\centering
\includegraphics[width=\textwidth]{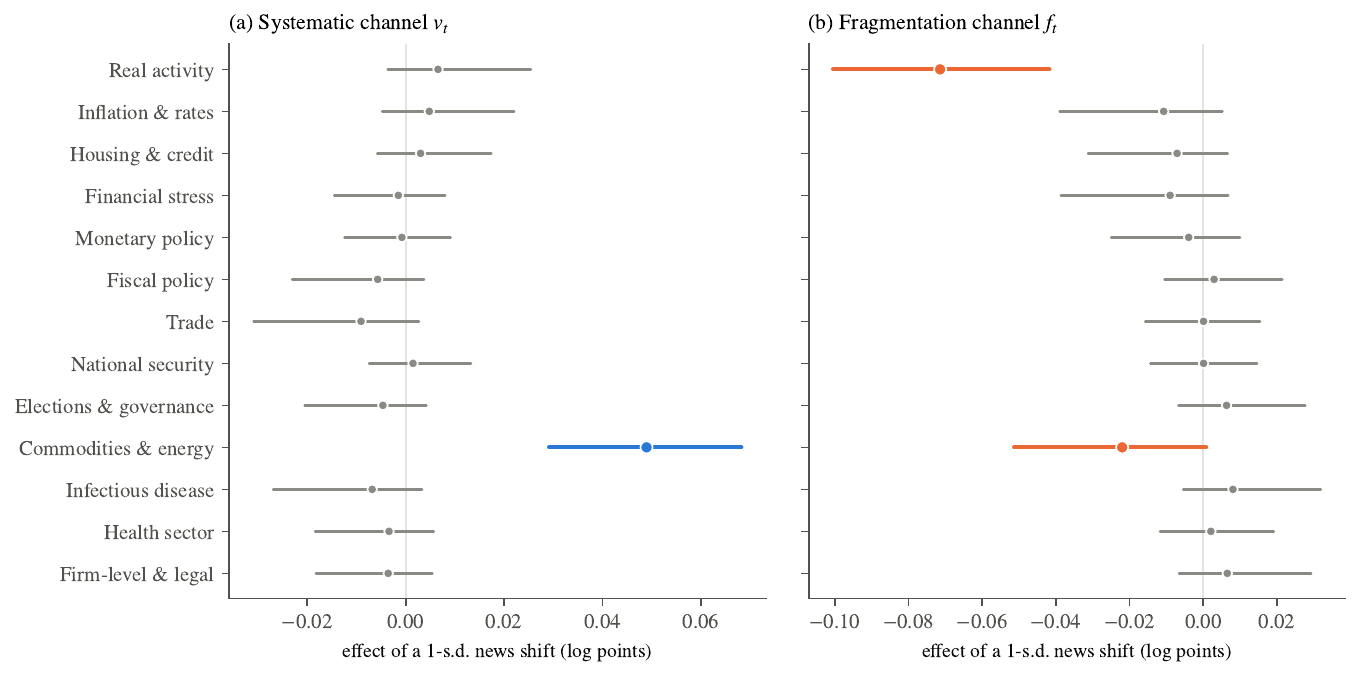}
\caption{News themes of the month and the two channels of chaos. (a) Effects on implied variance. (b) Effects on the fragmentation ratio. Dots are posterior means and bars are 90\% credible intervals, and colored effects have a posterior sign probability of at least 0.9.}\label{fig:news}
\end{figure}

To see which themes carry these associations, we estimate for each of $y_t$, $v_t$ and $f_t$ a regression on the $13$ theme regressors of the month in which the theme coefficients are shrunk. The regression for a series $z_t$ is
\begin{equation}\label{eq:themereg}
z_t=\beta_0+\bm\lambda^\top\bm q_{t-1}+\lambda_\iota\,\iota_t+\sum_{k=1}^K\beta_k\,x_{k,t}+\sigma\,\zeta_t+J_t,\qquad \zeta_t\sim\N(0,1),\qquad J_t\sim\mathrm{Exp}(\omega),
\end{equation}
where $\bm q_{t-1}$ contains $z_{t-1}$, the means of the last $3$ and of the last $12$ values of the series, and $v^{e}_{t-1}$, and where the exponential term $J_t$, which accommodates the right skewness of chaos and of implied variance, is omitted for the fragmentation ratio. The theme coefficients have the horseshoe prior~\cite{carvalho2010horseshoe}, that is,
\begin{equation}\label{eq:hs}
\beta_k\mid\tau_k,\tau\sim\N\big(0,\tau_k^2\tau^2\big),\qquad \tau_k\sim C^{+}(0,1),\qquad \tau\sim C^{+}(0,\tau_0),
\end{equation}
where $C^{+}$ denotes the half-Cauchy distribution and $\tau_0=0.1$. The horseshoe leaves large effects nearly unshrunk and shrinks small ones strongly. The remaining priors are $\N(0,100)$ for the intercept, $\N(0.5,0.25)$ truncated to $(-0.995,0.995)$ for the coefficient of $z_{t-1}$, $\N(0,1)$ for the other coefficients of the controls, inverse gamma with parameters $2$ and $0.02$ for $\sigma^2$, and gamma with shape $2$ and rate $0.2$, truncated to values above one, for $\omega$.

The posterior distribution is simulated by a Gibbs sampler which treats the bursts as missing data. Let $m_t$ denote the mean of $z_t-J_t$ in the regression~\eqref{eq:themereg}. Given the past, the density of $J_t$ is $\omega e^{-\omega J}$ on the nonnegative half-line, and the density of $z_t$ given $J_t=J$ is normal with mean $m_t+J$ and variance $\sigma^2$. By Bayes' rule and completion of the square, the burst given $z_t$ has a normal distribution truncated to the nonnegative half-line, namely
\begin{equation}\label{eq:jpost}
J_t\mid z_t\ \sim\ \N_{[0,\infty)}\big(z_t-m_t-\omega\sigma^2,\ \sigma^2\big).
\end{equation}
One sweep of the sampler consists of the following steps, in which $n$ is the number of months.
\begin{enumerate}[label=\normalfont(\arabic*)]
\item The bursts are drawn from the truncated normal distribution~\eqref{eq:jpost}.
\item The rate $\omega$ is drawn from the gamma distribution with shape $2+n$ and rate $0.2+\sum_tJ_t$, truncated to values above one.
\item The coefficients of the regression~\eqref{eq:themereg} are drawn from their Gaussian conditional distribution.
\item The local and global scales of the horseshoe prior~\eqref{eq:hs} are drawn by the auxiliary-variable scheme of Makalic and Schmidt~\cite{makalic2016simple}.
\item The variance $\sigma^2$ is drawn from the inverse-gamma distribution with shape $2+n/2$ and scale $0.02$ plus half of the sum of the squared residuals $z_t-J_t-m_t$.
\end{enumerate}
For the fragmentation ratio, steps~(1) and~(2) are omitted. We ran four chains of $2\times10^4$ iterations, discarded the first quarter and kept every tenth draw. The rank-normalized convergence statistic~\cite{vehtari2021rank} is at most $1.003$, and the bulk effective sample size is at least $1{,}296$.

Figure~\ref{fig:news} shows the posterior means and the $90\%$ credible intervals of the theme coefficients for implied variance and for the fragmentation ratio. Two themes stand out. Months with more news on real activity, that is, on output, employment and the sentiment of consumers and firms, have a lower fragmentation ratio, with a posterior mean of $-0.071$ and a $90\%$ interval from $-0.100$ to $-0.042$, and lower chaos, with a posterior mean of $-0.040$ and an interval from $-0.066$ to $-0.011$. Months with more news on commodities and energy have a higher implied variance, with a posterior mean of $0.049$ and an interval from $0.029$ to $0.068$, and a lower fragmentation ratio with posterior probability $0.920$. The least-squares regressions of Table~\ref{tab:news} with the themes added agree, with coefficients of real-activity news of $-0.077$ for the fragmentation ratio and of $-0.090$ for the co-movement term, and of commodity news of $0.046$ for implied variance, whose $t$-statistics are $-3.03$, $-2.71$ and $3.08$, respectively. When the news is about the economy as a whole or about commodity prices, returns are dominated by what stocks have in common. No theme is associated with fragmentation. The posterior distributions for elections, infectious disease and firm-level news lean toward a higher fragmentation ratio, but their intervals contain zero. The absence of a theme of fragmentation is the counterpart of low concentration, and it is consistent with fragmented chaos being the sum of many partial stories, each of which concerns a subset of firms.

\begin{table}[!ht]
\centering\footnotesize
\caption{Monte Carlo comparison of horseshoe and maximum-likelihood estimates of theme effects.}\label{tab:mchs}
\begin{tabular}{@{}lcccccc@{}}
\toprule
Theme & True & Posterior mean & RMSE, horseshoe & RMSE, likelihood & Coverage & Excludes zero\\
\midrule
Real activity & $0.000$ & $-0.001$ & 0.007 & 0.017 & 0.990 & 0.010\\
Inflation \& rates & $0.040$ & $0.030$ & 0.021 & 0.020 & 0.750 & 0.480\\
Housing \& credit & $0.000$ & $-0.000$ & 0.006 & 0.015 & 0.990 & 0.010\\
Financial stress & $-0.040$ & $-0.036$ & 0.021 & 0.021 & 0.820 & 0.620\\
Monetary policy & $0.000$ & $0.001$ & 0.008 & 0.019 & 0.990 & 0.010\\
Fiscal policy & $0.000$ & $0.000$ & 0.006 & 0.017 & 0.990 & 0.010\\
Trade & $0.030$ & $0.020$ & 0.017 & 0.016 & 0.740 & 0.370\\
National security & $0.000$ & $0.002$ & 0.006 & 0.016 & 1.000 & 0.000\\
Elections \& governance & $0.000$ & $0.000$ & 0.007 & 0.015 & 0.990 & 0.010\\
Commodities \& energy & $0.000$ & $0.001$ & 0.007 & 0.017 & 0.990 & 0.010\\
Infectious disease & $0.000$ & $-0.001$ & 0.007 & 0.019 & 1.000 & 0.000\\
Health sector & $0.000$ & $-0.001$ & 0.006 & 0.016 & 0.990 & 0.010\\
Firm-level \& legal & $0.000$ & $0.002$ & 0.007 & 0.015 & 0.970 & 0.030\\
\bottomrule
\end{tabular}
\end{table}

Two checks support this reading of the theme effects. The first is a Monte Carlo experiment on the horseshoe estimates in samples of our size. We simulated $100$ samples of chaos from a regression of the form~\eqref{eq:themereg} at its maximum-likelihood estimate, with the observed theme regressors, with the means of the last $3$ and of the last $12$ values computed on each simulated path, and with three nonzero theme effects, namely $0.04$ for inflation and rates, $-0.04$ for financial stress and $0.03$ for trade. Each sample was estimated by maximum likelihood and by the Gibbs sampler with $4{,}000$ iterations. Table~\ref{tab:mchs} reports, for every theme, the true effect, the mean over the samples of the posterior mean, the root mean squared error (RMSE) of the posterior mean and of the maximum-likelihood estimate, the coverage of the $90\%$ credible interval, and the share of the samples in which this interval excludes zero. For the ten themes without effect, the horseshoe reduces the RMSE from $0.016$ to $0.007$ on average, and the interval excludes zero in $0.010$ of the samples. For the three themes with an effect, the RMSE is nearly unchanged, the posterior means are shrunk toward zero, the coverage is between $0.740$ and $0.820$, and the interval excludes zero in $0.370$ to $0.620$ of the samples. An effect which the horseshoe reports as credible is therefore unlikely to be spurious, whereas the absence of a credible effect is weak evidence against effects of this size.

\begin{table}[!ht]
\centering\footnotesize
\caption{Sensitivity of theme effects to the global scale of the horseshoe prior.}\label{tab:sens}
\begin{tabular}{@{}lcccccc@{}}
\toprule
 & \multicolumn{2}{c}{$\tau_0=0.0025$} & \multicolumn{2}{c}{$\tau_0=0.1$} & \multicolumn{2}{c}{$\tau_0=1$}\\
\cmidrule(lr){2-3}\cmidrule(lr){4-5}\cmidrule(lr){6-7}
Theme & Mean & $\Pr(\beta_k>0)$ & Mean & $\Pr(\beta_k>0)$ & Mean & $\Pr(\beta_k>0)$\\
\midrule
\multicolumn{7}{@{}l}{\emph{Effects on $y_t$}}\\
\quad Real activity & $-0.038$ & 0.020 & $-0.040$ & 0.009 & $-0.040$ & 0.010\\
\quad Inflation \& rates & $0.022$ & 0.923 & $0.025$ & 0.953 & $0.025$ & 0.952\\
\multicolumn{7}{@{}l}{\emph{Effects on $v_t$}}\\
\quad Commodities \& energy & $0.051$ & 0.999 & $0.049$ & 1.000 & $0.049$ & 0.999\\
\multicolumn{7}{@{}l}{\emph{Effects on $f_t$}}\\
\quad Real activity & $-0.072$ & 0.000 & $-0.071$ & 0.000 & $-0.071$ & 0.000\\
\quad Commodities \& energy & $-0.018$ & 0.133 & $-0.022$ & 0.080 & $-0.022$ & 0.087\\
\bottomrule
\end{tabular}
\end{table}

The second check is the sensitivity of the theme effects to the prior. Table~\ref{tab:sens} reports the posterior mean and the posterior probability of a positive sign of the effects of news on real activity, on inflation and rates, and on commodities and energy, for three values of the global scale $\tau_0$ of the horseshoe prior~\eqref{eq:hs}. The value $0.0025$ follows the calibration of Piironen and Vehtari~\cite{piironen2017sparsity} for three relevant themes out of $13$, the value $0.1$ is the baseline, and the value $1$ is diffuse. The posterior summaries change little over this range, because the half-Cauchy prior lets the data determine the global shrinkage.

The news of the previous month has no robust predictive content for chaos. In Panel~B of Table~\ref{tab:news}, no coefficient of lagged intensity or lagged concentration is significant at the $5\%$ level. The Wald statistic of the $13$ lagged themes in the regression for chaos is $33.5$, which would have a $p$-value of $0.001$ under the chi-squared distribution and has a $p$-value of $0.174$ under the wild bootstrap. The end-of-month dispersion $y^{e}_{t-1}$ matters for this result. Without it and without the lags of the realized market variance among the controls, the Wald statistic is $27.6$, with a wild-bootstrap $p$-value of $0.284$, and adding $y^{e}_{t-1}$ alone, which enters with a coefficient of $0.162$ and a $t$-statistic of $4.60$, lowers the statistic to $22.0$, with a wild-bootstrap $p$-value of $0.466$. Lagged themes partly stand in for the level of dispersion at the end of the month, which monthly means do not capture. For implied variance, the lagged themes have no content either, with a wild-bootstrap $p$-value of $0.899$.

\section{Eras and Bursts}\label{sec:eras}

Sections~\ref{sec:faces} and~\ref{sec:news} studied chaos episode by episode and month by month. This section describes its dynamics. Figure~\ref{fig:annot} shows three features which a dynamic description must reproduce, namely persistence, with a first-order autocorrelation of $0.833$ for $y_t$, abrupt upward bursts which decay gradually, and long stretches at different levels, such as the late 1990s and the 2010s. Ataei et al.~\cite{ataei2021theory} dated the changes of level by change-point detection. Here, we add hidden regimes to the regression~\eqref{eq:themereg}, state what the resulting model implies for the tail of chaos, examine how strongly the data support its components, and evaluate its forecasts against benchmarks.

Let $S_t\in\{1,\dots,R\}$ be a hidden Markov regime. The model is
\begin{equation}\label{eq:modelA}
y_t=\alpha_{S_t}+\phi\,y_{t-1}+\lambda_v\,v^{e}_{t-1}+\lambda_\iota\,\iota_{t-1}+\bm\gamma_{S_t}^\top\vx_{t-1}+\sigma_{S_t}\zeta_t+J_t,\qquad \zeta_t\sim\N(0,1),\qquad J_t\mid S_t\sim\mathrm{Exp}(\omega_{S_t}),
\end{equation}
where the regimes are ordered by their intercepts, $\alpha_1<\dots<\alpha_R$, the end-of-month implied variance $v^{e}_{t-1}$ and the intensity $\iota_{t-1}$ are controls, and $\vx_{t-1}$ is the vector of the $13$ theme regressors of Definition~\ref{def:news}. The regimes describe eras, the exponential term $J_t\ge0$ describes bursts, and $\phi$ describes the persistence from month to month. In the HAR variant of the model, the means of the last $3$ and of the last $12$ values of $y_t$ are added to the regressors, which approximates long memory~\cite{corsi2009simple}. Conditionally on the regime and on the past, $y_t$ has the exponentially modified Gaussian distribution $\EMG(m,\sigma,\omega)$ with density
\begin{equation}\label{eq:emg}
p(y)=\omega\exp\Big\{-\omega(y-m)+\frac12\,\omega^2\sigma^2\Big\}\,\Phi\Big(\frac{y-m}{\sigma}-\omega\sigma\Big),
\end{equation}
where $\Phi$ is the standard normal distribution function, so that the monthly index in levels, $e^{y_t}$, has the modified lognormal power-law distribution~\cite{basu2015mlp}, to mixtures of which Ataei et al.~\cite{ataei2021theory} fitted the index. The regimes follow an ordered probit with transition probabilities which may depend on the news~\cite{diebold1994regime,filardo1994business}, namely
\begin{equation}\label{eq:tvtp}
\Pr(S_t=j\mid S_{t-1}=i)=\Phi\big(\kappa_{ij}-\bm\delta^\top\vx_{t-1}-\delta_\iota\,\iota_{t-1}\big)-\Phi\big(\kappa_{i,j-1}-\bm\delta^\top\vx_{t-1}-\delta_\iota\,\iota_{t-1}\big),
\end{equation}
with thresholds $-\infty=\kappa_{i0}<\kappa_{i1}<\dots<\kappa_{iR}=\infty$. A positive loading shifts probability toward the regimes of higher chaos from every starting regime. We require each regime to persist with probability at least $0.5$ at average news, since otherwise the likelihood is maximized by regimes which alternate from month to month and duplicate the bursts. The news loadings are pooled across regimes as $\bm\gamma_r=\bar{\bm\gamma}+\tilde{\bm\gamma}_r$, and the elements of $\bar{\bm\gamma}$, $\tilde{\bm\gamma}_r$ and $\bm\delta$ have the horseshoe prior~\eqref{eq:hs}. The next proposition states what the bursts imply for the upper tail of chaos, in the model without regressors and with constant transition probabilities.

\begin{proposition}[Bursts and tails]\label{prop:tail}
Let $y_t$ be the stationary solution of the model~\eqref{eq:modelA} with $R$ regimes, with $0\le\phi<1$, with $\lambda_v=\lambda_\iota=0$ and $\bm\gamma_r=\bm 0$, and with a constant transition matrix which is irreducible. Let $\bar\omega=\min_r\omega_r$. Then, the following hold.
\begin{enumerate}[label=\normalfont(\roman*)]
\item If a random variable $Y$ has the distribution $\EMG(m,\sigma,\omega)$, then $\E[e^{kY}]$ is finite if and only if $k<\omega$, and, as $x\to\infty$,
\begin{equation}\label{eq:tail1}
\Pr\big(e^{Y}>x\big)\sim e^{\omega m+\omega^2\sigma^2/2}\,x^{-\omega}.
\end{equation}
\item There exists a constant $C\in(0,\infty)$ such that, as $x\to\infty$,
\begin{equation}\label{eq:tail2}
\Pr\big(e^{y_t}>x\big)\sim C\,x^{-\bar\omega}.
\end{equation}
\item If $\phi>0$, if regime $r$ can follow in one step a regime whose tail index equals $\bar\omega$, and if $\omega_r\ne\bar\omega/\phi$, then there exists a constant $C_r\in(0,\infty)$ such that, as $x\to\infty$,
\begin{equation}\label{eq:tail3}
\Pr\big(e^{y_t}>x\mid S_t=r\big)\sim C_r\,x^{-\min(\omega_r,\,\bar\omega/\phi)}.
\end{equation}
\end{enumerate}
\end{proposition}

The proof is given in \ref{app:proofs}. Part~(i) is the power-law tail of the modified lognormal power-law distribution~\cite{basu2015mlp}, and part~(ii) shows that the stationary distribution of chaos inherits the tail of the most burst-prone regime. The present result adds part~(iii). The rate $\omega_r$ is the tail index of the one-step predictive distribution in regime $r$, but it is the tail index of chaos within regime $r$ only if it is smaller than $\bar\omega/\phi$. Otherwise, the tail within the regime is inherited, through the autoregressive term, from a preceding month in the most burst-prone regime. With bounded regressors, the solution of the model~\eqref{eq:modelA} differs from the solution without regressors by at most a constant, so that the exponents of parts~(ii) and~(iii) are unchanged and the tail probabilities are bounded above and below by constant multiples of the stated powers. With transition probabilities which depend on the news, we expect the same exponents, and the estimated transition loadings are close to zero.

The model is estimated in two ways. Maximum likelihood is computed by an expectation--conditional-maximization algorithm~\cite{dempster1977maximum,meng1993maximum}, which treats the regimes and the bursts as missing data and is fast enough for bootstrap tests and for repeated estimation out of sample. The algorithm takes the series $y_t$ and the regressors of the model~\eqref{eq:modelA} as input, with a constant transition matrix $\bm P=[P_{ij}]$. Write the mean of $y_t-J_t$ in regime $r$ as $m_{r,t}=\bm d_{r,t}^\top\bm\beta$, where $\bm d_{r,t}$ collects the regime indicators, $y_{t-1}$, the controls and, when the themes enter, the theme regressors interacted with the regime. Given $y_t$, the past and $S_t=r$, the burst has the truncated normal distribution~\eqref{eq:jpost} with $m_{r,t}$, $\sigma_r$ and $\omega_r$ in place of $m_t$, $\sigma$ and $\omega$, which separates a jump from a move to a higher regime. One iteration consists of the following steps.
\begin{enumerate}[label=\normalfont(\arabic*)]
\item The forward and backward recursions with the densities~\eqref{eq:emg} give the smoothed probabilities $\xi_{t,r}$ of the regimes and $\xi_{t,ij}$ of the pairs of consecutive regimes.
\item By the truncated normal distribution~\eqref{eq:jpost}, the conditional mean and variance of the burst are
\begin{equation*}
\bar J_{t,r}=a_{t,r}+\sigma_rh_{t,r},\qquad V_{t,r}=\sigma_r^2\Big(1-\frac{a_{t,r}}{\sigma_r}\,h_{t,r}-h_{t,r}^2\Big),
\end{equation*}
where $a_{t,r}=y_t-m_{r,t}-\omega_r\sigma_r^2$ and $h_{t,r}=\varphi(a_{t,r}/\sigma_r)/\Phi(a_{t,r}/\sigma_r)$, with $\varphi$ the standard normal density.
\item The tail indices, the variances and the transition probabilities are updated in closed form, namely
\begin{equation}\label{eq:cm}
\omega_r=\frac{\sum_t\xi_{t,r}}{\sum_t\xi_{t,r}\bar J_{t,r}},\qquad \sigma_r^2=\frac{\sum_t\xi_{t,r}\big\{(y_t-\bar J_{t,r}-m_{r,t})^2+V_{t,r}\big\}}{\sum_t\xi_{t,r}},\qquad P_{ij}=\frac{\sum_t\xi_{t,ij}}{\sum_t\sum_k\xi_{t,ik}}.
\end{equation}
\item The coefficients $\bm\beta$ are updated by weighted least squares of $y_t-\bar J_{t,r}$ on $\bm d_{r,t}$ with weights $\xi_{t,r}/\sigma_r^2$.
\end{enumerate}
The updates respect the restrictions $|\phi|\le0.995$, $1.01\le\omega_r\le50$ and $P_{ii}\ge0.5$, the last by a proportional rescaling of the off-diagonal elements of a row. Each step does not decrease the expected complete-data log-likelihood, so that the likelihood does not decrease~\cite{meng1993maximum}. The likelihood has several local maxima. We start from $30$ to $40$ classifications of the months, obtained from quantiles of $y_t$, of autoregressive residuals and of their absolute values, and from random persistent chains, run $60$ iterations from each, iterate the best runs until the relative change of the likelihood is below $10^{-9}$, and keep the best. Bootstrap replications use $12$ to $20$ starts, and out-of-sample estimation uses $10$ starts and the estimate of the previous year. Since a short window can drive the variance of a regime to zero, out-of-sample estimation maximizes a penalized likelihood~\cite{chen2008inference} with an inverse-gamma penalty with parameters $2$ and $0.02$ on each variance, which adds $0.04$ to the numerator and $6$ to the denominator of the variance update~\eqref{eq:cm}.

The posterior distribution is simulated by an extension of the Gibbs sampler of Section~\ref{sec:news}. Within each regime, the bursts, the tail index, the coefficients, the horseshoe scales and the variance are drawn as in the five steps of that sampler, with $m_{r,t}$, $\sigma_r$, $\omega_r$ and the number $n_r$ of months in regime $r$ in place of $m_t$, $\sigma$, $\omega$ and $n$, and with the coefficients restricted to ordered intercepts and to $|\phi|<0.995$ by rejection. The sweep adds the following steps.
\begin{enumerate}[label=\normalfont(\arabic*)]
\item Before the bursts, the regime path is drawn by forward filtering and backward sampling~\cite{chib1996calculating}, with the bursts integrated out by the densities~\eqref{eq:emg} and with the transition probabilities~\eqref{eq:tvtp}.
\item The thresholds of the ordered probit~\eqref{eq:tvtp} are drawn by a random-walk Metropolis--Hastings step with the latent normal random variables of the probit integrated out~\cite{cowles1996accelerating}.
\item The latent random variables of the probit are drawn from truncated normal distributions, and the transition loadings are drawn from their Gaussian conditional distribution~\cite{albert1993bayesian}, followed by their horseshoe scales.
\item The initial regime probabilities are drawn from their Dirichlet conditional distribution.
\end{enumerate}
The regimes and the bursts are thus drawn jointly, as are the thresholds and the latent random variables, so that the sweep is a partially collapsed sampler~\cite{vandyk2008partially}. The priors of Section~\ref{sec:news} apply within each regime, the thresholds have $\N(0,4)$ priors, and the global scale of the horseshoe prior is $\tau_0=0.1$ for the news loadings and $\tau_0=1$ for the transition loadings. The truncation of the tail indices to values above one makes the mean of the index in levels finite. With four chains of $2\times10^4$ iterations, of which the first quarter is discarded and every tenth draw is kept, the rank-normalized convergence statistic~\cite{vehtari2021rank} of the three-regime model is at most $1.014$ and the bulk effective sample size is at least $370$ for the intercepts, variances, tail indices and news loadings, and they are $1.023$ and $156$ for the transition thresholds, which mix most slowly.

\begin{table}[!ht]
\centering\footnotesize
\caption{Monte Carlo performance of the maximum-likelihood estimator.}\label{tab:mcecm}
\begin{tabular}{@{}lccccc@{}}
\toprule
Parameter & True & Mean & Bias & RMSE & Median\\
\midrule
$\alpha_1$ & $-6.090$ & $-6.184$ & $-0.094$ & 0.352 & $-6.210$\\
$\alpha_2$ & $-5.885$ & $-5.984$ & $-0.099$ & 0.334 & $-5.992$\\
$\alpha_3$ & $-5.615$ & $-5.706$ & $-0.092$ & 0.323 & $-5.724$\\
$\phi$ & $0.261$ & $0.249$ & $-0.012$ & 0.044 & $0.246$\\
$\sigma_1$ & $0.222$ & $0.204$ & $-0.018$ & 0.052 & $0.216$\\
$\sigma_2$ & $0.134$ & $0.132$ & $-0.003$ & 0.035 & $0.131$\\
$\sigma_3$ & $0.139$ & $0.133$ & $-0.006$ & 0.068 & $0.129$\\
$\omega_1$ & $12.300$ & $24.465$ & $12.165$ & 21.330 & $15.671$\\
$\omega_2$ & $5.007$ & $6.736$ & $1.730$ & 7.536 & $5.070$\\
$\omega_3$ & $2.269$ & $4.119$ & $1.850$ & 7.447 & $2.398$\\
$P_{11}$ & $0.984$ & $0.963$ & $-0.021$ & 0.068 & $0.982$\\
$P_{22}$ & $0.988$ & $0.972$ & $-0.016$ & 0.066 & $0.988$\\
$P_{33}$ & $0.968$ & $0.944$ & $-0.024$ & 0.075 & $0.966$\\
\bottomrule
\end{tabular}
\end{table}

A Monte Carlo experiment assesses the maximum-likelihood estimator in samples of our size. We simulated $200$ samples of $431$ months from the maximum-likelihood estimate of the three-regime model with regime-specific tail indices and without themes, with the controls at their observed values, and estimated the model on each sample by the expectation--conditional-maximization algorithm. Table~\ref{tab:mcecm} reports the true value, the mean, the bias, the root mean squared error (RMSE) and the median of the estimates. The autoregressive coefficient, the standard deviations and the persistence probabilities are recovered with small bias. The intercepts are imprecise individually, because they trade off against the autoregressive coefficient. The tail indices are the weakest part of the model. Their sampling distributions are strongly skewed to the right, because in some samples a regime contains few large bursts and its tail index drifts to the upper bound of $50$. Their medians, $15.671$, $5.070$ and $2.398$, are close to the true values $12.300$, $5.007$ and $2.269$, but their means are not. The modal smoothed regime coincides with the simulated regime in $0.922$ of the months on average and in more than $0.9$ of the months in $0.895$ of the samples.

\begin{table}[!ht]
\centering\footnotesize\setlength{\tabcolsep}{4.5pt}
\caption{Model selection for the monthly index.}\label{tab:select}
\begin{tabular}{@{}lcccccccc@{}}
\toprule
\multicolumn{9}{@{}l}{\emph{Panel A. Log-likelihood and BIC by number of regimes $R$}}\\
 & \multicolumn{2}{c}{$R=1$} & \multicolumn{2}{c}{$R=2$} & \multicolumn{2}{c}{$R=3$} & \multicolumn{2}{c}{$R=4$}\\
\cmidrule(lr){2-3}\cmidrule(lr){4-5}\cmidrule(lr){6-7}\cmidrule(lr){8-9}
 & $\log L$ & BIC & $\log L$ & BIC & $\log L$ & BIC & $\log L$ & BIC\\\midrule
AR(1), Gaussian & $-100.6$ & 231.5 & $-74.5$ & 209.6 & $-40.1$ & \textbf{183.4} & $-24.3$ & 206.4\\
AR(1), EMG & $-90.4$ & 217.1 & $-74.5$ & 215.7 & $-38.5$ & \textbf{186.1} & $-21.2$ & 206.1\\
HAR, Gaussian & $-67.4$ & 177.2 & $-38.1$ & \textbf{149.1} & $-23.0$ & 161.2 & $-7.2$ & 184.2\\
HAR, EMG & $-40.4$ & \textbf{129.4} & $-27.2$ & 133.2 & $-14.5$ & 150.4 & $-2.0$ & 179.9\\
AR(1), EMG, regime-specific $\omega_r$ & & & & & $-35.0$ & 191.3 & & \\
HAR, Student-$t$ & $-51.4$ & 151.3 & & & & & & \\
HAR, Gaussian, level-dependent variance & $-55.2$ & 158.9 & & & & & & \\
\midrule
\multicolumn{9}{@{}l}{\emph{Panel B. Parametric-bootstrap likelihood-ratio tests}}\\
Null $\to$ alternative & \multicolumn{2}{c}{Class} & LR & Replications & \multicolumn{2}{c}{Bootstrap $p$} & \multicolumn{2}{c}{95\% critical value}\\\midrule
$R=2\to R=3$ & \multicolumn{2}{c}{AR(1), EMG} & 72.1 & 99 & \multicolumn{2}{c}{$\le0.010$} & \multicolumn{2}{c}{14.6}\\
$R=3\to R=4$ & \multicolumn{2}{c}{AR(1), EMG} & 34.6 & 99 & \multicolumn{2}{c}{0.020} & \multicolumn{2}{c}{17.3}\\
Gaussian $\to$ EMG ($R=3$) & \multicolumn{2}{c}{AR(1)} & 3.3 & 499 & \multicolumn{2}{c}{0.038} & \multicolumn{2}{c}{2.8}\\
Common $\to$ regime-specific $\omega_r$ & \multicolumn{2}{c}{AR(1), EMG} & 6.9 & 499 & \multicolumn{2}{c}{0.058} & \multicolumn{2}{c}{7.6}\\
$R=1\to R=2$ & \multicolumn{2}{c}{HAR, EMG} & 26.5 & 99 & \multicolumn{2}{c}{$\le0.010$} & \multicolumn{2}{c}{13.3}\\
$R=1\to R=3$ & \multicolumn{2}{c}{HAR, EMG} & 51.8 & 99 & \multicolumn{2}{c}{$\le0.010$} & \multicolumn{2}{c}{24.8}\\
\midrule
\multicolumn{9}{@{}l}{\emph{Panel C. Exact local Whittle estimate $\hat d$ of $y_t$ under fitted models (observed $0.708$, 500 simulations)}}\\
Model & \multicolumn{2}{c}{Mean $\hat d$} & \multicolumn{2}{c}{5\%--95\%} & \multicolumn{2}{c}{$\Pr(\hat d\ge\hat d^{\mathrm{obs}})$} & & \\\midrule
AR(1), Gaussian & \multicolumn{2}{c}{0.520} & \multicolumn{2}{c}{0.361--0.661} & \multicolumn{2}{c}{0.004} & & \\
AR(1), EMG & \multicolumn{2}{c}{0.441} & \multicolumn{2}{c}{0.300--0.617} & \multicolumn{2}{c}{0.000} & & \\
Three-regime AR(1), EMG & \multicolumn{2}{c}{0.552} & \multicolumn{2}{c}{0.258--0.727} & \multicolumn{2}{c}{0.068} & & \\
HAR, Gaussian & \multicolumn{2}{c}{0.774} & \multicolumn{2}{c}{0.630--0.922} & \multicolumn{2}{c}{0.760} & & \\
HAR, EMG & \multicolumn{2}{c}{0.697} & \multicolumn{2}{c}{0.570--0.847} & \multicolumn{2}{c}{0.406} & & \\
\bottomrule
\end{tabular}
\end{table}

Table~\ref{tab:select} examines the support of the data for regimes, bursts and long memory, in the model without themes. Panel~A reports the log-likelihood and the Bayesian information criterion (BIC) for one to four regimes, for autoregressive and HAR dynamics, and for Gaussian and EMG innovations, with the smallest BIC of each row in bold. It adds two single-regime HAR models without bursts, one with Student-$t$ innovations and one with Gaussian innovations whose log variance is linear in $y_{t-1}$. Panel~B reports likelihood-ratio tests whose $p$-values and critical values are obtained by a parametric bootstrap~\cite{mclachlan1987bootstrapping}, in which both models are estimated again from multiple starts in every replication. The tests use $99$ replications, and $499$ for the two tests of the innovation distribution, whose $p$-values are close to the conventional levels. The bootstrap is needed because the number of regimes is not identified under the null hypothesis~\cite{hansen1992likelihood} and because the Gaussian model lies on the boundary of the parameter space of the EMG model. Panel~C compares the exact local Whittle estimate~\cite{shimotsu2005exact} of the memory parameter of $y_t$, which is $0.708$, with its distribution in $500$ samples simulated from fitted models.

Three conclusions follow. First, the innovations of chaos are right-skewed and fat-tailed, and the exponential burst captures both. In the single-regime HAR class, the log-likelihood rises from $-67.4$ for Gaussian innovations to $-51.4$ for Student-$t$ innovations and to $-40.4$ for EMG innovations, which have the same number of parameters as the Student-$t$. The gain of the Student-$t$ over the Gaussian reflects the fat tail, and the further gain of the EMG reflects the skewness. The residuals of the Gaussian HAR model have a skewness of $1.057$ and an excess kurtosis of $4.113$. Second, regimes and long memory are substitutes for the slow movement of chaos, and the data do not discriminate sharply between them. With autoregressive dynamics, the BIC selects three regimes and the bootstrap tests reject two regimes against three. With HAR dynamics, the single-regime EMG model has the smallest BIC of all models of the table, although the bootstrap tests still reject one regime. The three-regime autoregressive model reproduces the observed memory estimate only in the upper tail of its simulated distribution, whereas the HAR models reproduce it without difficulty. Third, within the three-regime model the evidence for bursts and for regime-specific tail indices is weaker than the evidence for regimes. The likelihood-ratio statistics are $3.3$ and $6.9$, with bootstrap $p$-values of $0.038$ and $0.058$, respectively. We therefore use the three-regime model as a description of the eras of chaos and not as the selected model, and we carry the single-regime HAR model with bursts alongside it in the forecast comparison.

\begin{table}[!ht]
\centering\footnotesize
\caption{Unit-root, stationarity and long-memory statistics.}\label{tab:persist}
\begin{tabular}{@{}llccccc@{}}
\toprule
Frequency & Series & Observations & ADF & DF-GLS & KPSS & $\hat d$\\
\midrule
Monthly & $y$ (log FCIX) & 432 & $-2.974$ (0.037) & $-2.953$ (0.003) & 0.306 ($\ge0.100$) & 0.708 (0.070)\\
Monthly & $v$ (log implied variance) & 432 & $-3.788$ (0.003) & $-3.017$ (0.003) & 0.115 ($\ge0.100$) & 0.614 (0.070)\\
Monthly & $f$ (fragmentation ratio) & 432 & $-2.998$ (0.035) & $-1.611$ (0.105) & 0.358 (0.095) & 0.682 (0.070)\\
Daily & $y$ (log FCIX) & 9067 & $-5.253$ ($<0.001$) & $-5.224$ ($<0.001$) & 1.580 ($\le0.010$) & 0.526 (0.026)\\
Daily & $v$ (log implied variance) & 9063 & $-6.435$ ($<0.001$) & $-6.359$ ($<0.001$) & 0.405 (0.075) & 0.765 (0.026)\\
Daily & $f$ (fragmentation ratio) & 9063 & $-6.660$ ($<0.001$) & $-6.283$ ($<0.001$) & 1.576 ($\le0.010$) & 0.470 (0.026)\\
\bottomrule
\end{tabular}
\end{table}

The persistence which regimes and long memory both describe is documented in Table~\ref{tab:persist}. It reports, for chaos, implied variance and the fragmentation ratio at the monthly and at the daily frequency, the augmented Dickey--Fuller (ADF) statistic, the DF-GLS statistic~\cite{elliott1996efficient} and the statistic of Kwiatkowski et al.~\cite{kwiatkowski1992testing} (KPSS), with $p$-values in parentheses, and the exact local Whittle estimate $\hat d$ of the memory parameter~\cite{shimotsu2010exact}, with its standard error in parentheses. The memory estimate uses the first $\lfloor n^{0.65}\rfloor$ frequencies of a series of length $n$. The ADF test rejects a unit root for every series, and the DF-GLS test does so for every series except the monthly fragmentation ratio. The KPSS test does not reject stationarity of monthly chaos and implied variance at the $10\%$ level, and it rejects stationarity of daily chaos and of the daily fragmentation ratio at the $1\%$ level. The memory estimates lie between $0.470$ and $0.765$. The three series are thus persistent, with the slowly decaying dependence which regimes and long memory both produce~\cite{diebold2001long}.

\begin{table}[!ht]
\centering\footnotesize
\caption{Posterior summary of the three-regime model.}\label{tab:era}
\begin{tabular}{@{}lcccccc@{}}
\toprule
 & \multicolumn{2}{c}{Calm} & \multicolumn{2}{c}{Turbulent} & \multicolumn{2}{c}{Highly chaotic}\\
\cmidrule(lr){2-3}\cmidrule(lr){4-5}\cmidrule(lr){6-7}
Parameter & Mean & 90\% interval & Mean & 90\% interval & Mean & 90\% interval\\
\midrule
Intercept $\alpha_r$ & $-6.707$ & $[-7.400,\,-5.969]$ & $-6.427$ & $[-7.077,\,-5.735]$ & $-5.929$ & $[-6.575,\,-5.281]$\\
Gaussian standard deviation $\sigma_r$ & $0.198$ & $[0.158,\,0.234]$ & $0.144$ & $[0.110,\,0.185]$ & $0.178$ & $[0.080,\,0.306]$\\
Tail index $\omega_r$ & $10.587$ & $[5.375,\,21.381]$ & $6.230$ & $[4.299,\,9.959]$ & $5.543$ & $[2.373,\,15.185]$\\
Mean burst $1/\omega_r$ & $0.113$ & $[0.047,\,0.186]$ & $0.173$ & $[0.100,\,0.233]$ & $0.251$ & $[0.066,\,0.421]$\\
Burst share of innovation variance & $0.261$ & $[0.041,\,0.566]$ & $0.579$ & $[0.234,\,0.800]$ & $0.611$ & $[0.052,\,0.956]$\\
Cross-sectional standard deviation (\%) & $1.741$ & $[1.695,\,1.790]$ & $2.145$ & $[2.076,\,2.213]$ & $3.060$ & $[2.812,\,3.317]$\\
Persistence $P_{rr}$ at average news & $0.984$ & $[0.964,\,0.997]$ & $0.985$ & $[0.963,\,0.998]$ & $0.945$ & $[0.869,\,0.990]$\\
Expected duration in months (median) & $73$ & $[27,\,351]$ & $82$ & $[27,\,468]$ & $22$ & $[8,\,98]$\\
Ergodic probability & $0.329$ & $[0.067,\,0.692]$ & $0.533$ & $[0.203,\,0.859]$ & $0.138$ & $[0.014,\,0.384]$\\
Share of months (smoothed probabilities) & $0.442$ &  & $0.421$ &  & $0.137$ & \\
\midrule
Autoregressive coefficient $\phi$ & $0.186$ & $[0.102,\,0.276]$ & & & & \\
End-of-month implied variance, $\lambda_{v}$ & $0.145$ & $[0.110,\,0.180]$ & & & & \\
News intensity, $\lambda_{\iota}$ & $0.002$ & $[-0.030,\,0.035]$ & & & & \\
Smallest tail index $\min_r\omega_r$ & $4.084$ & $[2.373,\,6.438]$ & & & & \\
\bottomrule
\end{tabular}
\end{table}

\begin{figure}[!ht]
\centering
\includegraphics[width=\textwidth]{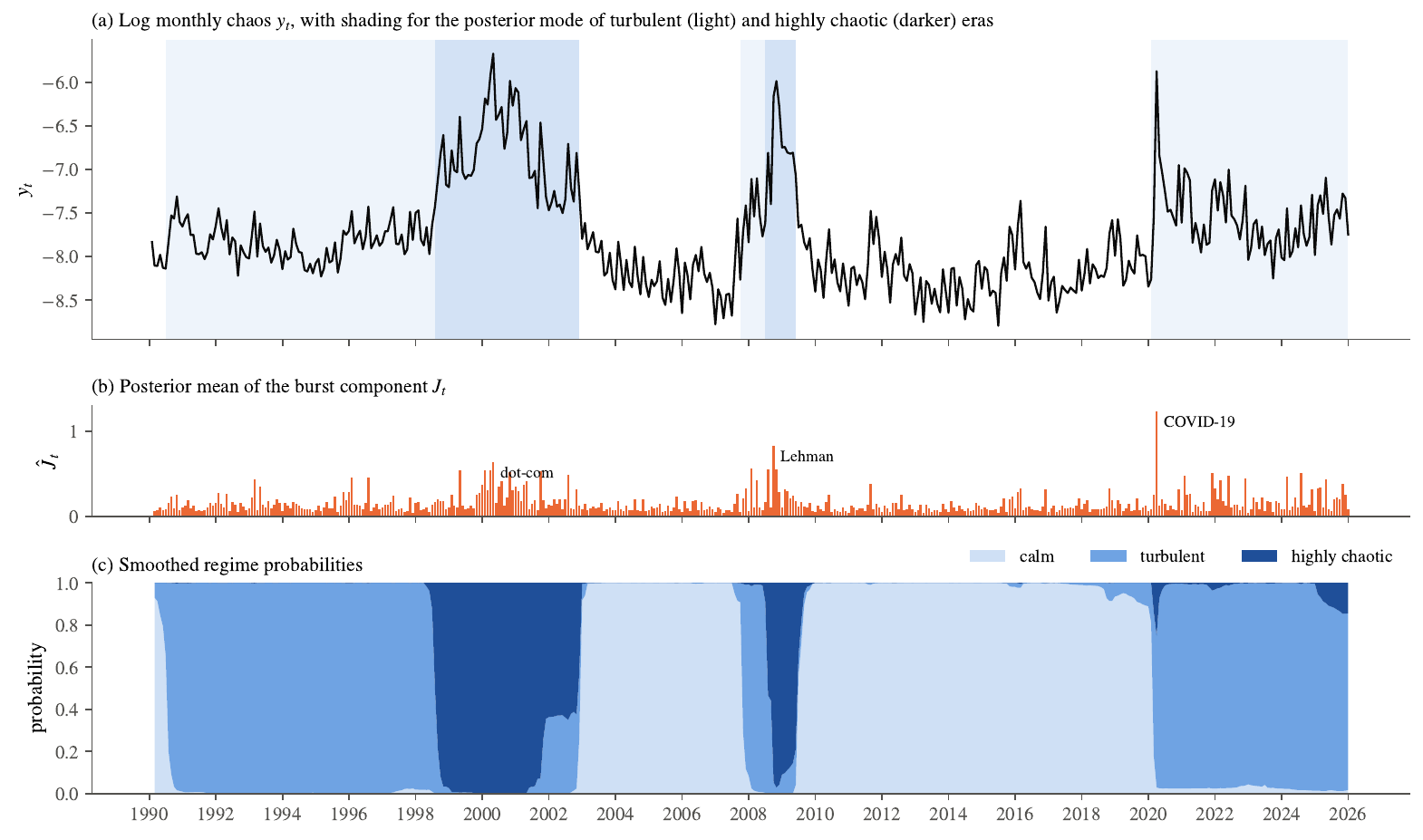}
\caption{Eras and bursts of chaos. (a) Monthly index $y_t$ with the modal regime. (b) Posterior mean of the burst $J_t$. (c) Smoothed regime probabilities.}\label{fig:eras}
\end{figure}

Table~\ref{tab:era} reports posterior means and $90\%$ credible intervals for the three-regime model with themes, by regime in its upper part and for the common parameters in its lower part, and Figure~\ref{fig:eras} shows the resulting chronology. The regimes are long eras. The calm regime covers December 2002 to September 2007 and June 2009 to January 2020, apart from five months in 1990, and the daily cross-sectional standard deviation of returns at its level is $1.741\%$. The turbulent regime covers most of the 1990s, the nine months before July 2008, and the whole period since February 2020, with a cross-sectional standard deviation of $2.145\%$. The highly chaotic regime is visited twice, from August 1998 to November 2002 and from July 2008 to May 2009, with a cross-sectional standard deviation of $3.060\%$. The posterior medians of the expected durations are $73$, $82$ and $22$ months, respectively.

\begin{table}[!ht]
\centering\footnotesize
\caption{Three-regime model by form of the index, universe and sample.}\label{tab:robustreg}
\begin{tabular}{@{}lcccccccc@{}}
\toprule
 & & \multicolumn{3}{c}{Tail index $\omega_r$} & \multicolumn{3}{c}{Persistence $P_{rr}$} & Regime\\
\cmidrule(lr){3-5}\cmidrule(lr){6-8}
Variant & Months & calm & turbulent & high & calm & turbulent & high & agreement\\
\midrule
Top 1,000, logarithmic form (baseline) & 432 & 12.300 & 5.007 & 2.269 & 0.984 & 0.988 & 0.968 & 1.000\\
Top 500, logarithmic form & 432 & 8.665 & 4.605 & 2.015 & 0.984 & 0.988 & 0.968 & 0.993\\
S\&P 500 constituents, logarithmic form & 432 & 5.525 & 15.075 & 3.395 & 0.992 & 0.987 & 0.968 & 0.735\\
Top 1,000, multiplicative form & 432 & 12.127 & 4.856 & 2.250 & 0.984 & 0.988 & 0.968 & 1.000\\
Baseline, 1990--2007 & 216 & 8.103 & 9.683 & 2.751 & 0.984 & 0.990 & 0.981 & 0.986\\
Baseline, 2008--2025 & 216 & 38.976 & 7.824 & 3.584 & 0.987 & 0.981 & 0.988 & 0.405\\
\bottomrule
\end{tabular}
\end{table}

The chronology is a description relative to the sample. Table~\ref{tab:robustreg} reports the maximum-likelihood estimates of the tail indices and of the persistence probabilities of the three-regime model without themes for the forms of the index and the universes of Table~\ref{tab:robust} and for the two halves of the sample, together with the share of the months whose modal regime agrees with the baseline. The chronology is unchanged for the multiplicative form and for the $500$ largest stocks, and it agrees in $0.735$ of the months for the constituents of the S\&P~500. Estimated on 1990--2007 alone, the regimes agree with the baseline in $0.986$ of the months of that period, whereas estimated on 2008--2025 alone they agree in $0.405$ of the months. The second half of the sample contains no level comparable to that of 1998--2002, and the regimes re-center on its own levels. The eras are thus not a set of fixed thresholds.

The two visits to the highly chaotic regime differ in the terms of the chaos accounting, which a classification by level cannot show. From August 1998 to November 2002, the fragmentation ratio averaged $1.452$, above its mean of $1.128$ over the sample, and the chaos-implied correlation and the realized co-movement ratio averaged $0.204$ and $0.155$, below their means of $0.255$ and $0.191$, respectively. From July 2008 to May 2009, the fragmentation ratio averaged $0.693$, and the chaos-implied correlation, the realized co-movement ratio and COR1M averaged $0.345$, $0.388$ and $0.585$, respectively, the last against a mean of $0.379$. The first visit was an era of high dispersion with ordinary co-movement, and the second an era in which stocks moved together.

\begin{table}[!ht]
\centering\footnotesize
\caption{Regime selection for implied variance and the fragmentation ratio.}\label{tab:twin}
\begin{tabular}{@{}lcccccccc@{}}
\toprule
 & \multicolumn{2}{c}{$R=1$} & \multicolumn{2}{c}{$R=2$} & \multicolumn{2}{c}{$R=3$} & \multicolumn{2}{c}{$R=4$}\\
\cmidrule(lr){2-3}\cmidrule(lr){4-5}\cmidrule(lr){6-7}\cmidrule(lr){8-9}
Series and model & $\log L$ & BIC & $\log L$ & BIC & $\log L$ & BIC & $\log L$ & BIC\\
\midrule
$v_t$, AR(1), Gaussian & $-41.6$ & 113.6 & $-8.5$ & \textbf{77.6} & $-4.8$ & 112.7 & $1.4$ & 154.9\\
$v_t$, AR(1), EMG & $-13.7$ & \textbf{63.8} & $-4.2$ & 75.2 & $-0.7$ & 110.7 & $2.9$ & 157.9\\
$v_t$, HAR, Gaussian & $-39.6$ & 121.7 & $-2.0$ & \textbf{76.7} & $2.6$ & 110.0 & & \\
$v_t$, HAR, EMG & $-7.2$ & \textbf{62.9} & $1.9$ & 75.0 & $8.3$ & 104.7 & & \\
$f_t$, AR(1), Gaussian & $-133.8$ & 298.0 & $-116.4$ & \textbf{293.4} & $-97.6$ & 298.3 & $-83.6$ & 324.9\\
$f_t$, AR(1), EMG & $-133.8$ & 304.1 & $-116.4$ & \textbf{299.5} & $-97.5$ & 304.3 & $-83.6$ & 331.0\\
$f_t$, HAR, Gaussian & $-98.4$ & \textbf{239.2} & $-83.4$ & 239.7 & $-75.1$ & 265.4 & $-69.4$ & 308.6\\
\bottomrule
\end{tabular}
\end{table}

Table~\ref{tab:twin} repeats Panel~A of Table~\ref{tab:select} for implied variance and for the fragmentation ratio, with the smallest BIC of each row in bold. For implied variance, the single-regime model with bursts has the smallest BIC with autoregressive and with HAR dynamics. Given its own value at the end of the previous month, implied variance behaves as a persistent series with bursts and without eras. For the fragmentation ratio, the BIC selects two Gaussian regimes with autoregressive dynamics, the second of which is occupied in the $28$ months from March 1999 to June 2001 and in no other month, and it does not discriminate between one and two regimes with HAR dynamics. The fragmentation ratio is thus episodic, which is why Section~\ref{sec:faces} studies it through episodes, and its one long episode lies within the first visit to the highly chaotic regime.

The bursts separate jumps from changes of era (see Figure~\ref{fig:eras}(b)). The three largest posterior mean bursts are $1.233$ in March 2020, $0.833$ in September 2008 and $0.635$ in April 2000. The peak of March 2020 is read by the model as a burst within the turbulent regime, and the posterior probability of the highly chaotic regime in that month is $0.251$. This reading depends on the regime-specific tail indices, since with a common tail index the maximum-likelihood fit assigns that month to the highly chaotic regime. The shape of chaos differs across regimes, although imprecisely. Bursts account for $0.261$ of the innovation variance in the calm regime and for about $0.6$ in the other two, and the posterior probability that the share is larger in the turbulent than in the calm regime is $0.893$.

The tail index of chaos is poorly determined, and most of the evidence places it between two and four. The maximum-likelihood estimates of the tail indices are $12.300$, $5.007$ and $2.269$ in the calm, turbulent and highly chaotic regime, with $\phi=0.261$, so that $\bar\omega/\phi=8.702$, and the highly chaotic regime is followed by the calm regime with positive probability. By Proposition~\ref{prop:tail}(iii), the tail index of chaos within the calm regime is then $8.702$ and not $12.300$. The posterior distribution of the smallest tail index has a median of $3.826$ and a $90\%$ interval from $2.373$ to $6.438$, against a prior interval from $1.437$ to $10.850$. A direct estimate which uses no model agrees. The estimator of Hill~\cite{hill1975simple} applied to the monthly index in levels gives $2.465$ with the largest $5\%$ and $2.056$ with the largest $10\%$ of the observations, and applied to the daily index it gives $3.100$ with the largest $1\%$ and $2.503$ with the largest $5\%$. Moments of chaos beyond the second may therefore not exist.

\begin{table}[!ht]
\centering\scriptsize
\caption{One-month-ahead forecasts of the monthly index, 2005--2025.}\label{tab:oos}
\begin{tabular}{@{}lcccccccccc@{}}
\toprule
Model & MSE & LS & LS$^{-}$ & CRPS & DM & Brier & AUC & Skewness & Berkowitz & MCS\\
\midrule
\multicolumn{11}{@{}l}{\emph{Gaussian benchmarks (least squares and elastic net)}}\\
\quad Random walk & $0.130$ & $-0.441$ & $-0.385$ & $0.202$ & $-2.06$ & $0.077$ & $0.691$ & $0.411$ & 0.000 & 0.001\\
\quad AR(1) & $0.118$ & $-0.386$ & $-0.323$ & $0.191$ & $-1.09$ & $0.076$ & $0.629$ & $0.610$ & 0.000 & 0.018\\
\quad HAR & $0.102$ & $-0.324$ & $-0.212$ & $0.171$ &  & $0.074$ & $0.692$ & $0.864$ & 0.031 & 0.772\\
\quad HAR with controls & $0.101$ & $-0.320$ & $-0.215$ & $0.172$ & $0.28$ & $0.074$ & $0.716$ & $0.732$ & 0.009 & 0.746\\
\quad HAR with controls and news (OLS) & $0.106$ & $-0.371$ & $-0.261$ & $0.178$ & $-1.16$ & $0.073$ & $0.730$ & $0.380$ & 0.000 & 0.389\\
\quad HAR with controls and news (elastic net) & $0.103$ & $-0.343$ & $-0.237$ & $0.174$ & $-0.50$ & $0.073$ & $0.737$ & $0.464$ & 0.000 & 0.607\\
\multicolumn{11}{@{}l}{\emph{Fat-tailed and heteroskedastic benchmarks (maximum likelihood)}}\\
\quad HAR with controls, Student-$t$ & $0.101$ & $-0.253$ & $-0.214$ & $0.171$ & $0.98$ & $0.074$ & $0.714$ & $0.535$ & 0.008 & 0.772\\
\quad HAR with controls, level-dependent variance & $0.100$ & $-0.315$ & $-0.217$ & $0.170$ & $0.37$ & $0.076$ & $0.702$ & $0.693$ & 0.000 & 0.772\\
\multicolumn{11}{@{}l}{\emph{Burst and regime models (maximum likelihood)}}\\
\quad Three regimes, Gaussian & $0.111$ & $-0.353$ & $-0.326$ & $0.180$ & $-0.29$ & $0.075$ & $0.646$ & $0.614$ & 0.000 & 0.607\\
\quad Three regimes, bursts & $0.105$ & $-0.284$ & $-0.257$ & $0.177$ & $0.41$ & $0.074$ & $0.686$ & $0.011$ & 0.001 & 0.772\\
\quad Three regimes, bursts, regime $\omega_r$ & $0.104$ & $-0.259$ & $-0.231$ & $0.173$ & $0.65$ & $0.073$ & $0.699$ & $-0.167$ & 0.001 & 0.772\\
\quad Three regimes, bursts, news & $0.124$ & $-0.375$ & $-0.350$ & $0.191$ & $-0.51$ & $0.076$ & $0.591$ & $0.071$ & 0.000 & 0.323\\
\quad HAR, bursts & $0.101$ & $-0.231$ & $-0.197$ & $0.169$ & $1.08$ & $0.074$ & $0.693$ & $-0.102$ & 0.000 & 0.772\\
\quad HAR, three regimes, bursts & $0.106$ & $-0.249$ & $-0.212$ & $0.172$ & $0.86$ & $0.074$ & $0.681$ & $-0.130$ & 0.000 & 0.772\\
\multicolumn{11}{@{}l}{\emph{Burst and regime models (Bayesian)}}\\
\quad HAR, bursts & $0.100$ & $-0.213$ & $-0.179$ & $0.168$ & $1.29$ & $0.073$ & $0.700$ & $0.049$ & 0.008 & 0.907\\
\quad HAR, bursts, shrunk news & $\mathbf{0.099}$ & $\mathbf{-0.205}$ & $\mathbf{-0.172}$ & $\mathbf{0.168}$ & $1.31$ & $0.073$ & $0.702$ & $0.003$ & 0.044 & 1.000\\
\quad Three regimes, bursts & $0.102$ & $-0.223$ & $-0.201$ & $0.173$ & $0.99$ & $0.072$ & $0.697$ & $0.035$ & 0.092 & 0.860\\
\quad Three regimes, bursts, shrunk news & $0.106$ & $-0.208$ & $-0.182$ & $0.172$ & $1.19$ & $\mathbf{0.071}$ & $\mathbf{0.788}$ & $0.084$ & 0.264 & 0.921\\
\bottomrule
\end{tabular}
\end{table}

We finally ask what regimes, bursts and news contribute to forecasts. We forecast $y_t$ one month ahead for the $252$ months from January 2005 to December 2025. Every model is estimated each January on all data up to the preceding December, its parameters are held fixed within the year, and the regime probabilities are updated monthly. All regressors are standardized with the moments of the estimation window. For the Bayesian models, the predictive densities of $200$ retained draws are averaged. Table~\ref{tab:oos} compares $18$ models in four groups. The Gaussian benchmarks are the random walk, the first-order autoregression, the HAR model, the HAR model with the controls of the model~\eqref{eq:modelA}, and the latter with the themes, estimated by least squares or by the elastic net~\cite{zou2005regularization}. The second group replaces the Gaussian innovations of the HAR model with controls by Student-$t$ innovations, or lets the log variance depend linearly on $y_{t-1}$. The third and fourth groups contain the burst and regime models, estimated by maximum likelihood and by the Gibbs sampler, respectively. The columns report the mean squared error (MSE), the mean log score in nats over all months (LS) and without March 2020 (LS$^{-}$), the mean continuous ranked probability score (CRPS)~\cite{gneiting2007strictly}, and the Diebold--Mariano statistic~\cite{diebold1995comparing} of the log score against the HAR model (DM), which is positive when a model is better. Since most models nest the HAR model, this statistic is a descriptive measure and not a formal test~\cite{clark2007approximately}. The next two columns evaluate the forecast of a surge, defined as a monthly change of $y_t$ above the 95th percentile of the changes in the estimation window, by the Brier score and by the area under the receiver operating characteristic curve (AUC). There are $21$ surges. The last three columns report the skewness of the normal scores of the probability integral transforms, the $p$-value of the test of Berkowitz~\cite{berkowitz2001testing} for their calibration, and the $p$-value of the model confidence set~\cite{hansen2011model} for the log score. The best value of a column is in bold.

\begin{figure}[!ht]
\centering
\includegraphics[width=\textwidth]{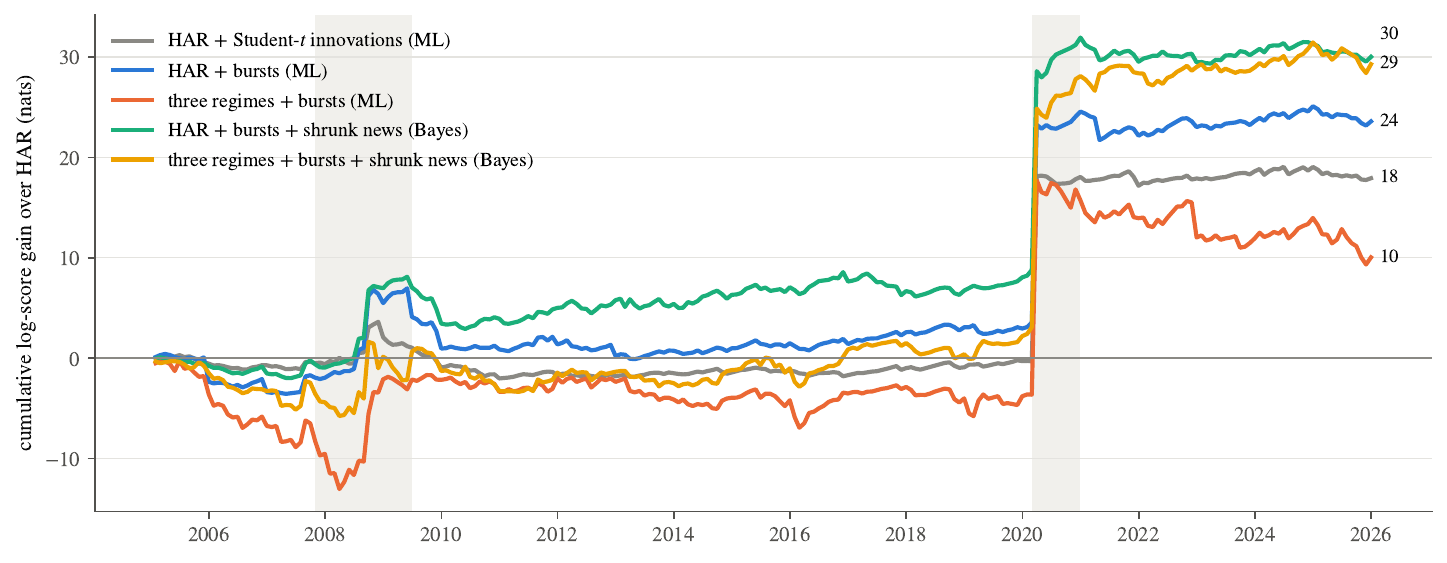}
\caption{Cumulative log-score gain over the HAR model, 2005--2025. Lines show selected models of Table~\ref{tab:oos}, and the shaded bands mark 2007--2009 and 2020.}\label{fig:oos}
\end{figure}

Three conclusions follow from Table~\ref{tab:oos} and Figure~\ref{fig:oos}. First, the density forecasts are separated by their tails, and by one month. The HAR model with controls has a mean log score of $-0.320$, the same model with Student-$t$ innovations has $-0.253$, and the HAR model with bursts has $-0.231$. Over the $252$ months, the burst model gains $22.4$ nats on the Gaussian model, of which $17.8$ are earned in March 2020, when the logarithms of the predictive densities at the realized value are $-26.6$, $-10.0$ and $-8.7$ for the three models. The Student-$t$ model obtains $16.7$ of the $22.4$ nats, that is, three quarters, and $16.6$ of them in March 2020. Without that month, the three models have mean log scores of $-0.215$, $-0.214$ and $-0.197$. The gain of bursts over a symmetric fat tail is $5.7$ nats, spread over the sample, with a Diebold--Mariano statistic of $0.82$. It stems from the skewness, which the Student-$t$ model leaves in its probability integral transforms, with a skewness of $0.535$ against $-0.102$ for the burst model. Second, no model is distinguishable from the HAR model in point or density forecasts by the usual criteria. The largest Diebold--Mariano statistic against the HAR model is $1.31$, and the model confidence set at the $10\%$ level excludes only the random walk and the autoregression. Estimation uncertainty matters for the regime models, whose Bayesian versions have better log scores than their maximum-likelihood versions, and unrestricted themes worsen every model, whereas shrunk themes change the scores little. Third, the forecasts of surges are weak for all models. The base rate of a surge is $21/252=0.083$, and the Brier score of the constant forecast at this rate is $0.076$. The best model, the Bayesian three-regime model with shrunk themes, attains $0.071$, and its AUC of $0.788$ exceeds the $0.692$ of the HAR model. A moving-block bootstrap gives a $90\%$ interval from $0.017$ to $0.173$ for the difference of the two AUC values. With $21$ surges and $18$ models, we regard this difference as suggestive.

\section{Conclusion}\label{sec:conclusion}

This paper studied the financial chaos index as a measure of what the VIX does not register. It first showed that the tensor index reduces to the ratio of the arithmetic to the harmonic mean of the gross returns when the consensus weights all assets equally, gave the exact factor by which the two differ otherwise, and adopted the closed form as the definition of the index, which makes the index a function of the cross-section of a single day. The index and its logarithmic and simple-return companions were shown to be averages of pairwise disagreements, which differ in the weight that they give to extreme pairs, and the logarithmic form was shown to estimate the cross-sectional variance of returns with an error which is small in most months and is dominated by the collapse of single stocks in a few. The paper then derived an accounting which splits the chaos of every month, exactly and in observed terms, into implied variance, a wedge between realized and implied market variance, the realized co-movement of stocks and a form gap. Applied to large U.S.\ stocks over 1990--2025, the accounting showed that between a third and a half of the episodes of high chaos are fragmented, that fragmented episodes are of the same order of magnitude as synchronized ones and occur throughout the sample, and that they consist of falling co-movement in about half of the cases and of turmoil of the broad market in excess of implied variance in most of the others. Intense news was found to accompany chaos of both kinds, and concentrated news to accompany higher implied variance and higher co-movement with no more dispersion, whereas lagged news had no robust predictive content. Finally, a regime-switching model with exponential bursts described calm, turbulent and highly chaotic eras, of which the two highly chaotic ones differed in co-movement, and the bursts were found to amount to right-skewed and fat-tailed innovations whose forecast gains over a long-memory benchmark stem mostly from one month. The motivation of the paper was that the VIX describes one portfolio. The accounting states how much of the turmoil of the cross-section that description leaves out, namely more than half of the variance of excess chaos, and what the remainder consists of in any given month.

Three directions for future work stand out. The first is an index which is robust to single stocks. The pairwise representation admits kernels which grow more slowly than the square, such as the absolute difference of two returns, for which a central limit theorem requires a finite second moment only, and the open problem is to carry the chaos accounting over to such a kernel, for which the exact split into market variance and co-movement is lost. The second is the wedge. With the weights of the index in place of equal weights, the realized side of the accounting refers to the same portfolio as the VIX, and the wedge then consists of the variance risk premium and of the surprise of the month. Separating these two parts, and the breadth of the turmoil, requires the weights of the constituents and a model of expected variance. The third is anticipation. Lagged news does not predict chaos, and whether the prices of options on individual stocks, which enter the implied correlation index, anticipate months of falling co-movement is an open question whose answer would show whether fragmented chaos can be hedged before it occurs.

\section*{Declaration of Competing Interest}

The authors declare no competing interests.

\section*{Data Availability}

The daily stock returns were obtained from CRSP through Wharton Research Data Services. The derived monthly series, the code and the result files which reproduce all tables and figures are available from the authors.

\appendix
\section{Proofs}\label{app:proofs}

This appendix contains the proofs of Theorems~\ref{thm:closed}--\ref{thm:dci} and of Propositions~\ref{prop:weighted}--\ref{prop:tail}, in the order of the main text, together with one auxiliary lemma.

\begin{proof}[Proof of Theorem~\ref{thm:closed}]
The three parts are proved in order, and the subscript $d$ is dropped.
\begin{enumerate}[label=\normalfont(\roman*)]
\item By Lemma~\ref{lem:scale}, the largest eigenvalue of the consensus slice is
\begin{equation*}
\lambda_{\max}\big(\tilde\mA\big)=\frac{\big(\tilde\vx^\top(1/\vr)\big)\big(\tilde\vy^\top\vr\big)\big(\tilde\vy^\top\tilde\vx\big)}{\|\tilde\vx\|^2\,\|\tilde\vy\|^2}.
\end{equation*}
Since $\tilde\vx=\bar x(\one+\va)$ and $\tilde\vy=\bar y(\one+\vb)$, where $\va$ and $\vb$ have mean zero, the five factors are
\begin{align*}
\tilde\vx^\top(1/\vr)&=\frac{N\bar x}{\HM}\big\{1+\HM\cov(\va,1/\vr)\big\}, & \tilde\vy^\top\vr&=N\bar y\,\AM\Big\{1+\frac{\cov(\vb,\vr)}{\AM}\Big\},\\
\tilde\vy^\top\tilde\vx&=N\bar x\bar y\,\big\{1+\cov(\va,\vb)\big\}, & \|\tilde\vx\|^2\|\tilde\vy\|^2&=N^2\bar x^2\bar y^2\,\big\{1+\var(\va)\big\}\big\{1+\var(\vb)\big\}.
\end{align*}
Substitution gives $\lambda_{\max}(\tilde\mA)=N(\AM/\HM)\,\Theta$, and Definition~\ref{def:tensorindex} then gives the formula~\eqref{eq:theta}.
\item If $\va=\vb=\bm 0$, every covariance and variance in $\Theta$ vanishes, so that $\Theta=1$, and the formula~\eqref{eq:theta} reduces to the closed form~\eqref{eq:fcix}.
\item If all assets earn the same return, then $\AM=\HM$, and the vectors $\vr$ and $1/\vr$ are constant, so that their covariances with $\va$ and $\vb$ vanish. With $s_a$ and $s_b$ the cross-sectional standard deviations of $\va$ and $\vb$, the Cauchy--Schwarz inequality gives
\begin{equation*}
\Theta=\frac{1+\cov(\va,\vb)}{(1+s_a^2)(1+s_b^2)}\le\frac{1+s_as_b}{(1+s_a^2)(1+s_b^2)}\le1,
\end{equation*}
where the last inequality holds because
\begin{equation*}
(1+s_a^2)(1+s_b^2)-(1+s_as_b)=(s_a-s_b)^2+s_as_b+s_a^2s_b^2\ge0,
\end{equation*}
with equality if and only if $s_a=s_b=0$. Hence, $\Fcp=N(\Theta-1)/(N-1)\le0$, with equality if and only if $\va=\vb=\bm 0$.\qedhere
\end{enumerate}
\end{proof}

\begin{proof}[Proof of Theorem~\ref{thm:ustat}]
The five parts are proved in order. We write $\tilde\ell_i=\ell_i-\bar\ell$ and $\mu_k=N^{-1}\sum_{i=1}^N\tilde\ell_i^{\,k}$, so that $\mu_2=\Fl$ and $\mathcal K=\mu_4/\mu_2^2$.
\begin{enumerate}[label=\normalfont(\roman*)]
\item By Definition~\ref{def:pcm}, the sum of all elements of the comparison matrix is
\begin{equation*}
\sum_{i,j=1}^Na_{ij}=\Big(\sum_{j=1}^Nr_j\Big)\Big(\sum_{i=1}^N\frac1{r_i}\Big)=N^2\,\frac{\AM}{\HM},
\end{equation*}
and its diagonal contributes $N$. Grouping the pair $(i,j)$ with the pair $(j,i)$ therefore gives
\begin{equation*}
\sum_{i\ne j}(a_{ij}-1)=N^2\Big(\frac{\AM}{\HM}-1\Big)=N(N-1)\,\F,\qquad a_{ij}+a_{ji}-2=2\big\{\cosh(\ell_i-\ell_j)-1\big\},
\end{equation*}
and division by $N(N-1)$ proves the first identity. For any real numbers $u_1,\dots,u_N$ with mean $\bar u$,
\begin{equation*}
\sum_{i<j}\frac{(u_i-u_j)^2}{2}=\frac14\sum_{i,j=1}^N(u_i-u_j)^2=\frac N2\sum_{i=1}^N(u_i-\bar u)^2,
\end{equation*}
and division by $N(N-1)/2$ proves the second and the third identity, with $u_i=\ell_i$ and $u_i=R_i$, respectively.
\item By the first display of part~(i), $\mathcal S=N\AM/\HM$, so that $(\mathcal S-N)/(N-1)=\F$. If $\bm w$ is the Perron vector of a positive matrix $\mA$, then $(\mA\bm w)_i=\lambda_{\max}w_i$ for every $i$, and averaging over $i$ gives the representation of Saaty~\cite{saaty1977scaling},
\begin{equation*}
\lambda_{\max}=\frac1N\sum_{i,j=1}^Na_{ij}\,\frac{w_j}{w_i}.
\end{equation*}
The quantity $\mathcal S$ is this expression at $\bm w=\one$.
\item The matrix $\mL=\one\tilde{\bm\ell}^\top-\tilde{\bm\ell}\one^\top$ is skew-symmetric, and since $\one^\top\tilde{\bm\ell}=0$,
\begin{equation*}
\mL^\top\mL=\|\tilde{\bm\ell}\|^2\,\one\one^\top+N\,\tilde{\bm\ell}\tilde{\bm\ell}^\top
\end{equation*}
is the sum of two matrices of rank at most one with orthogonal ranges, each with the eigenvalue $N\|\tilde{\bm\ell}\|^2$. Hence, the two largest singular values of $\mL$ equal $\sqrt N\|\tilde{\bm\ell}\|$, and the squared Frobenius norm is their sum of squares, namely
\begin{equation*}
\|\mL\|_F^2=2N\|\tilde{\bm\ell}\|^2=2N^2\,\Fl.
\end{equation*}
\item By parts~(i) and~(iii), the ratio $\AM/\HM$ is the average of $\cosh L_{ij}$ over all pairs $(i,j)$, and $\Fl$ is half of the average of $L_{ij}^2$. The power series of the hyperbolic cosine therefore gives
\begin{equation*}
\frac{N-1}{N}\,\F-\Fl=\frac1{N^2}\sum_{i,j=1}^N\Big(\cosh L_{ij}-1-\frac12L_{ij}^2\Big)=\sum_{k\ge2}\frac{M_{2k}}{(2k)!},
\end{equation*}
and every term is nonnegative. Expanding $(\tilde\ell_i-\tilde\ell_j)^4$ and using $\sum_i\tilde\ell_i=0$ gives
\begin{equation*}
\sum_{i,j=1}^NL_{ij}^4=2N^2\mu_4+6N^2\mu_2^2,\qquad\text{that is,}\qquad \frac{M_4}{24}=\frac{\mu_2^2\,(\mathcal K+3)}{12},
\end{equation*}
which is the term with $k=2$.
\item Since $r_i=\AM(1+e_i)$ and the deviations $e_i$ have mean zero, the identity $1/(1+e)=1-e+e^2/(1+e)$ gives
\begin{equation*}
\frac{\AM}{\HM}-1=\frac1N\sum_{i=1}^N\frac{1}{1+e_i}-1=\frac1N\sum_{i=1}^N\frac{e_i^2}{1+e_i},
\end{equation*}
and $\CSV$ is the cross-sectional variance of $\AM(1+e_i)$, which is $\AM^2m_2$. For the logarithmic form, $\ell_i=\log\AM+\log(1+e_i)$, and for $|e|\le1/2$ the expansions
\begin{equation*}
\log(1+e)=e-\frac{e^2}{2}+\frac{e^3}{3}-\frac{e^4}{4}+O\big(|e|^5\big),\qquad \{\log(1+e)\}^2=e^2-e^3+\frac{11}{12}\,e^4+O\big(|e|^5\big)
\end{equation*}
hold with remainders bounded by a constant multiple of $|e|^5$. Averaging over the assets gives
\begin{equation*}
\frac1N\sum_{i=1}^N\log(1+e_i)=-\frac{m_2}{2}+\frac{m_3}{3}-\frac{m_4}{4}+O\Big(\max_i|e_i|^5\Big),\qquad \frac1N\sum_{i=1}^N\{\log(1+e_i)\}^2=m_2-m_3+\frac{11}{12}\,m_4+O\Big(\max_i|e_i|^5\Big),
\end{equation*}
and the square of the first mean is $m_2^2/4$ up to a remainder of the same order. The logarithmic form is the second mean minus the square of the first, which is the expansion~\eqref{eq:logexp}.\qedhere
\end{enumerate}
\end{proof}

\begin{proof}[Proof of Theorem~\ref{thm:clt}]
The four parts are proved in order, and all moments are conditional on $\mathcal G$.
\begin{enumerate}[label=\normalfont(\roman*)]
\item By Theorem~\ref{thm:ustat}(i), the rescaled index $N\Fl_d/(N-1)$ is a $U$-statistic of order two with the kernel $h(x,x')=(x-x')^2/2$, whose expectation for two independent draws of the cross-section is the variance $D_d$. A $U$-statistic is unbiased for the expectation of its kernel for every $N\ge2$.
\item Let $\mu$ and $\mu^\ast_4$ be the mean and the fourth central moment of the cross-section, so that $\mathcal K^\ast_d=\mu^\ast_4/D_d^2$. The kernel has a finite second moment if and only if $\mu^\ast_4$ is finite, and its projection is
\begin{equation*}
h_1(x)=\E\,h(x,\ell_{1,d})=\frac12\big\{(x-\mu)^2+D_d\big\},\qquad 4\Var h_1(\ell_{1,d})=\mu^\ast_4-D_d^2.
\end{equation*}
By the central limit theorem for $U$-statistics of Hoeffding~\cite{hoeffding1948class}, $\sqrt N\{N\Fl_d/(N-1)-D_d\}$ converges in distribution to $\N(0,\mu^\ast_4-D_d^2)$. The factor $N/(N-1)$ changes the statistic by a term of order $N^{-1}$, and the delta method for the logarithm, with derivative $1/D_d$, gives the limit~\eqref{eq:clt}.
\item With $\tilde\theta_{i,d}=\theta_{i,d}-\bar\theta_d$ and $\bar\varepsilon_d$ the cross-sectional mean of the shocks, expanding the square in the definition of $\Fl_d$ gives
\begin{equation}\label{eq:psidec}
\Fl_d-D_d=\frac1N\sum_{i=1}^N\big\{\varepsilon_{i,d}^2-h_{i,d}+2\tilde\theta_{i,d}\,\varepsilon_{i,d}\big\}-\bar\varepsilon_d^{\,2}.
\end{equation}
The summands are independent with mean zero and uniformly bounded variances, and $\E\,\bar\varepsilon_d^{\,2}=O(N^{-1})$, so that $\Fl_d-D_d=O_p(N^{-1/2})$. By the definition of $U_t$, the monthly index~\eqref{eq:monthly} is
\begin{equation*}
y_t=\log\bigg(\frac{1}{n_t}\sum_{d\in t}D_d\bigg)+\log(1+U_t),
\end{equation*}
and $\log(1+U_t)=U_t+O_p(N^{-1})$ because $U_t=O_p(N^{-1/2})$. The indexes of different days are independent given $\mathcal G$, so that the variance of $U_t$ is the ratio of $\sum_{d\in t}\Var(\Fl_d)$ to $(\sum_{d\in t}D_d)^2$, and $\Var(\Fl_d)=D_d^2(\mathcal K^\circ_d-1)/N$ by the definition of the effective kurtosis. The inequality follows by replacing each $\mathcal K^\circ_d$ by the largest.
\item Let $A_d$ denote the average on the right-hand side of the decomposition~\eqref{eq:psidec}, and let $\chi_{i,d}$ denote its summands. Since $s_{i,d}$ is the sum of $\tilde\theta_{i,d}^{\,2}$ and $h_{i,d}$, a direct computation gives
\begin{equation*}
\Var(\chi_{i,d})=\E\,\varepsilon_{i,d}^4+4\tilde\theta_{i,d}\,\E\,\varepsilon_{i,d}^3+4\tilde\theta_{i,d}^{\,2}h_{i,d}-h_{i,d}^2=\E\big(\tilde\theta_{i,d}+\varepsilon_{i,d}\big)^4-s_{i,d}^2.
\end{equation*}
By the independence across assets and the bounded fourth moments, it follows that
\begin{equation*}
N\Var(A_d)=\bar\mu_{4,d}-\frac1N\sum_{i=1}^Ns_{i,d}^2,\qquad \Var\big(\bar\varepsilon_d^{\,2}\big)=O\big(N^{-2}\big),\qquad \Cov\big(A_d,\bar\varepsilon_d^{\,2}\big)=\frac{1}{N^3}\sum_{i=1}^N\Cov\big(\chi_{i,d},\varepsilon_{i,d}^2\big)=O\big(N^{-2}\big).
\end{equation*}
Hence, the decomposition~\eqref{eq:psidec} gives
\begin{equation*}
N\Var(\Fl_d)=\bar\mu_{4,d}-\frac1N\sum_{i=1}^Ns_{i,d}^2+O\big(N^{-1}\big)=\bar\mu_{4,d}-D_d^2-\var(\bm s_d)+O\big(N^{-1}\big),
\end{equation*}
and division by $D_d^2$ gives the relation~\eqref{eq:keff}.\qedhere
\end{enumerate}
\end{proof}

\begin{proof}[Proof of Theorem~\ref{thm:dci}]
The three parts are proved in order, and the subscript $d$ is dropped in parts~(i) and~(iii).
\begin{enumerate}[label=\normalfont(\roman*)]
\item Since $\mSigma$ is positive semidefinite with largest eigenvalue $\lambda_{\max}(\mSigma)$,
\begin{equation*}
0\le\one^\top\mSigma\one\le N\lambda_{\max}(\mSigma)\le N\tr\mSigma,
\end{equation*}
which gives $0\le\bar\rho\le1$. With the centering matrix $\bm M=\bm I-N^{-1}\one\one^\top$, the dispersion is the quadratic form $\CSV=N^{-1}\vR^\top\bm M\vR$, whose conditional expectation is
\begin{equation*}
\E\big[\CSV\mid\mathcal F\big]=\frac1N\tr(\bm M\mSigma)+\frac1N\,\bm\mu^\top\bm M\bm\mu=\bar\sigma^2-\sigma^2_{\mathrm{EW}}+V,
\end{equation*}
and $\bar\sigma^2-\sigma^2_{\mathrm{EW}}=\bar\sigma^2(1-\bar\rho)$.
\item The average squared return of day $d$ is $N^{-1}\sum_iR_{i,d}^2=\bar R_d^{\,2}+\CSV_d$. By the definition~\eqref{eq:rhotilde}, the ratio of $\sum_{d\in t}\CSV_d$ to $\sum_{d\in t}\bar R_d^{\,2}$ therefore equals $(1-\tilde\rho_t)/\tilde\rho_t$, and taking logarithms gives the identity~\eqref{eq:realized}.
\item If $\bm\mu=\bm 0$, then $\E[\bar R^{\,2}\mid\mathcal F]$ is the variance $\one^\top\mSigma\one/N^2$ of the equal-weighted return, and $\E[N^{-1}\sum_iR_i^2\mid\mathcal F]=\tr\mSigma/N$.\qedhere
\end{enumerate}
\end{proof}

\begin{proof}[Proof of Proposition~\ref{prop:weighted}]
The three parts are proved in order.
\begin{enumerate}[label=\normalfont(\roman*)]
\item Expanding the square and taking conditional expectations gives
\begin{equation*}
\CSV_\varpi=\sum_{i=1}^N\varpi_iR_i^2-R_\varpi^2,\qquad \E\big[\CSV_\varpi\mid\mathcal F\big]=\sum_{i=1}^N\varpi_i\big(\Sigma_{ii}+\mu_i^2\big)-\bm\varpi^\top\mSigma\bm\varpi-\big(\bm\varpi^\top\bm\mu\big)^2,
\end{equation*}
and the second expression is the stated one. By the convexity of the square, the variance $\bm\varpi^\top\mSigma\bm\varpi$ of a convex combination of the returns is at most the convex combination $\bar\sigma^2_\varpi$ of their variances, which gives $\bar\rho_\varpi\le1$.
\item The first identity is the expansion of the square used in part~(i). For the second, the sum over the pairs with $i<j$ is half of the sum over all pairs, namely
\begin{equation*}
\sum_{i<j}\varpi_i\varpi_j\,(R_i-R_j)^2=\frac12\sum_{i,j=1}^N\varpi_i\varpi_j\,\big(R_i^2-2R_iR_j+R_j^2\big)=\sum_{i=1}^N\varpi_iR_i^2-R_\varpi^2.
\end{equation*}
\item Since $a_{ij}=r_j/r_i$, the double sum factorizes.\qedhere
\end{enumerate}
\end{proof}

\begin{proof}[Proof of Proposition~\ref{prop:alias}]
The two parts are proved in order.
\begin{enumerate}[label=\normalfont(\roman*)]
\item Since $\bm\varsigma_t=\bm{\mathcal E}_t-\iota_t\one$, the systematic part of $z_t$ is
\begin{equation*}
\beta_0+\beta_\iota\,\iota_t+\bm\eta^\top\bm\varsigma_t=\beta_0+\bm\eta^\top\bm{\mathcal E}_t+\big(\beta_\iota-\one^\top\bm\eta\big)\iota_t,
\end{equation*}
which is a linear function of $\bm{\mathcal E}_t$ for all values of $\iota_t$ and $\bm\varsigma_t$ if and only if $\beta_\iota=\one^\top\bm\eta$.
\item Under the stated assumptions, the covariance matrix of $\bm{\mathcal E}_t$ is $\sigma_\iota^2\one\one^\top+\mSigma_\varsigma$, and its covariance with $z_t$ is $\beta_\iota\sigma_\iota^2\one$. By the Sherman--Morrison formula,
\begin{equation*}
\big(\mSigma_\varsigma+\sigma_\iota^2\one\one^\top\big)^{-1}\one=\frac{\mSigma_\varsigma^{-1}\one}{1+\varrho},
\end{equation*}
which gives the coefficients~\eqref{eq:alias}. For two categories whose log shares have standard deviations $d_1>d_2$ and correlation $\rho_{12}$, the vector $\mSigma_\varsigma^{-1}\one$ is proportional to $(d_2^2-\rho_{12}d_1d_2,\,d_1^2-\rho_{12}d_1d_2)^\top$, so that the first coefficient has the sign opposite to that of $\beta_\iota$ whenever $\rho_{12}>d_2/d_1$.\qedhere
\end{enumerate}
\end{proof}

The proof of Proposition~\ref{prop:tail} uses the following lemma on the sum of a random variable with an exponential tail and an independent random variable with a lighter tail.

\begin{lemma}[Convolution with an exponential tail]\label{lem:conv}
Let $X$ and $Z$ be independent random variables, let $\theta>0$ and $K\ge0$, and suppose that $e^{\theta x}\Pr(X>x)\to K$ as $x\to\infty$ and that $\E[e^{\theta Z}]<\infty$. Then, as $x\to\infty$,
\begin{equation*}
e^{\theta x}\Pr(X+Z>x)\to K\,\E\big[e^{\theta Z}\big].
\end{equation*}
\end{lemma}

\begin{proof}
The function $q(x)=e^{\theta x}\Pr(X>x)$ converges as $x\to\infty$ and is at most $e^{\theta x}$, so that it is bounded on the real line by a constant $M$. By independence,
\begin{equation*}
e^{\theta x}\Pr(X+Z>x)=\E\big[e^{\theta Z}\,q(x-Z)\big].
\end{equation*}
The integrand converges to $Ke^{\theta Z}$ as $x\to\infty$ and is dominated by $Me^{\theta Z}$, which is integrable, and the claim follows by dominated convergence.
\end{proof}

\begin{proof}[Proof of Proposition~\ref{prop:tail}]
The three parts are proved in order.
\begin{enumerate}[label=\normalfont(\roman*)]
\item Write $Y=m+\sigma Z_0+J$, where $Z_0\sim\N(0,1)$ and $J\sim\mathrm{Exp}(\omega)$ are independent. Then, for $k<\omega$,
\begin{equation*}
\E\big[e^{kY}\big]=e^{km+k^2\sigma^2/2}\,\frac{\omega}{\omega-k},
\end{equation*}
and the expectation is infinite for $k\ge\omega$. Since $e^{\omega x}\Pr(J>x)=1$ for $x\ge0$, Lemma~\ref{lem:conv} with $X=J$, $Z=m+\sigma Z_0$ and $\theta=\omega$ gives $e^{\omega x}\Pr(Y>x)\to e^{\omega m+\omega^2\sigma^2/2}$, which is the tail~\eqref{eq:tail1} at $x$ replaced by $\log x$.
\item The stationary solution and its decomposition into the burst of month $t$ and a remainder are
\begin{equation*}
y_t=\sum_{k\ge0}\phi^k\big(\alpha_{S_{t-k}}+\sigma_{S_{t-k}}\zeta_{t-k}+J_{t-k}\big),\qquad y_t=J_t+W_t,\qquad W_t=\alpha_{S_t}+\sigma_{S_t}\zeta_t+\phi\,y_{t-1}.
\end{equation*} Given the regime path, the innovations $\zeta_t$ and the bursts $J_t$ are independent, and for $0\le s<\bar\omega$,
\begin{equation*}
\E\big[e^{sy_t}\mid(S_u)\big]\le\exp\Big\{\frac{s\max_r|\alpha_r|}{1-\phi}+\frac{s^2\max_r\sigma_r^2}{2(1-\phi^2)}\Big\}\prod_{k\ge0}\frac{\bar\omega}{\bar\omega-s\phi^k}<\infty,
\end{equation*}
since the moment generating function $\omega/(\omega-a)$ of an exponential random variable with rate $\omega\ge\bar\omega$ is at most $\bar\omega/(\bar\omega-a)$ for $0\le a<\bar\omega$. The bound does not depend on the regime path. Applied to $y_{t-1}$ with $s\phi$ in place of $s$, it shows that $\E[e^{sW_t}\mid S_t=r]$ is finite for every $s$ with $s\phi<\bar\omega$.

Consider a regime $r$ with $\omega_r\phi<\bar\omega$, which includes every regime with $\omega_r=\bar\omega$. Given $S_t=r$, the burst $J_t$ is independent of $W_t$, and Lemma~\ref{lem:conv} with $X=J_t$, $Z=W_t$ and $\theta=\omega_r$ gives
\begin{equation}\label{eq:tailr}
e^{\omega_rx}\Pr(y_t>x\mid S_t=r)\to C_r=\E\big[e^{\omega_rW_t}\mid S_t=r\big]\in(0,\infty).
\end{equation}
Consider next a regime $r$ with $\omega_r\phi\ge\bar\omega$, and fix $s$ with $\bar\omega<s<\bar\omega/\phi$, so that $s<\omega_r$. By the Markov inequality and the independence of $J_t$ and $W_t$ given $S_t=r$,
\begin{equation}\label{eq:tailo}
\Pr(y_t>x\mid S_t=r)\le e^{-sx}\,\E\big[e^{sJ_t}\mid S_t=r\big]\,\E\big[e^{sW_t}\mid S_t=r\big]=o\big(e^{-\bar\omega x}\big).
\end{equation}
The stationary probabilities $\pi_r$ of an irreducible chain are positive. Summing the limits~\eqref{eq:tailr} and the bounds~\eqref{eq:tailo} over the regimes gives $e^{\bar\omega x}\Pr(y_t>x)\to C$, where $C$ is the sum of $\pi_rC_r$ over the regimes with $\omega_r=\bar\omega$, which is the tail~\eqref{eq:tail2} at $x$ replaced by $\log x$.
\item If $\omega_r<\bar\omega/\phi$, the claim is the limit~\eqref{eq:tailr}. Let $\omega_r>\bar\omega/\phi$. Given $S_{t-1}$, the value $y_{t-1}$ is independent of $S_t$, since the transition matrix is constant. Hence,
\begin{equation*}
\Pr(y_{t-1}>x\mid S_t=r)=\sum_{q=1}^R\Pr(S_{t-1}=q\mid S_t=r)\,\Pr(y_{t-1}>x\mid S_{t-1}=q),
\end{equation*}
and by the limits~\eqref{eq:tailr} and the bounds~\eqref{eq:tailo}, $e^{\bar\omega x}\Pr(y_{t-1}>x\mid S_t=r)\to K_r$, where $K_r$ is the sum of $\Pr(S_{t-1}=q\mid S_t=r)\,C_q$ over the regimes $q$ with $\omega_q=\bar\omega$. The constant $K_r$ is positive because regime $r$ can follow such a regime in one step. With $\theta=\bar\omega/\phi$, it follows that $e^{\theta x}\Pr(\phi\,y_{t-1}>x\mid S_t=r)\to K_r$. Given $S_t=r$, the random variables $\phi\,y_{t-1}$, $\alpha_r+\sigma_r\zeta_t$ and $J_t$ are independent, the second has finite exponential moments of every order, and $\E[e^{\theta J_t}\mid S_t=r]=\omega_r/(\omega_r-\theta)$ is finite. Two applications of Lemma~\ref{lem:conv} give
\begin{equation*}
e^{\theta x}\Pr(y_t>x\mid S_t=r)\to K_r\,e^{\theta\alpha_r+\theta^2\sigma_r^2/2}\,\frac{\omega_r}{\omega_r-\theta},
\end{equation*}
which is the tail~\eqref{eq:tail3} with the exponent $\bar\omega/\phi$.\qedhere
\end{enumerate}
\end{proof}

\end{document}